\documentclass[english,11pt,a4paper]{article}
\pdfoutput=1
\usepackage{jheppub}
\usepackage[T1]{fontenc}
\usepackage[latin9]{inputenc}
\usepackage{amsmath}
\usepackage{amsthm}
\usepackage{graphicx}

\makeatletter

\providecommand{\tabularnewline}{\\}

\usepackage{slashed}
\usepackage{tabularx}
\usepackage{tikz}
\usetikzlibrary{shapes,arrows}
\allowdisplaybreaks

\makeatother

\usepackage{babel}
\begin{document}
\title{Dark matter electromagnetic dipoles: the WIMP expectation}
\author{Thomas Hambye and Xun-Jie Xu}
\affiliation{Service de Physique Th\'{e}orique, Universit\'{e} Libre de Bruxelles, Boulevard du Triomphe, CP225, 1050 Brussels, Belgium}
\date{\today}
\abstract{
We perform a systematic study of the electric and magnetic dipole moments of dark matter (DM) that are induced 
at the one-loop level when DM experiences four-fermion interactions with Standard Model (SM) charged fermions. 
Related to their loop nature these moments can largely depend on the UV completion at the origin of the four-fermion operators. We illustrate this property
by considering explicitly two simple ways to generate these operators,
from $t$- or $s$-channel tree-level exchange.
Fixing the strength of these interactions from the DM relic density constraint, we obtain in particular a magnetic moment that, depending on the interaction considered, lies typically between $10^{-20}$ to $10^{-23}$ ecm or identically vanishes. 
These non-vanishing values 
induce, via photon exchange, DM-nucleus scattering cross sections
that could be probed by current or near future direct detection experiments. 
}
\maketitle



\section{Introduction}

\noindent Weakly interacting massive particles (WIMPs) are among
the best motivated dark matter (DM) candidates. 
As is well known,  DM particles annihilating into lighter particles with coupling strength of order unity undergo a non-relativistic freeze-out in the primordial thermal bath of the Universe, leaving a relic density of the order of the observed one if the DM mass is roughly around the electroweak scale. This ``WIMP miracle'' has triggered
vast experimental effort in DM searches---see \cite{Jungman:1995df,Bertone:2004pz,Feng:2010gw,Arcadi:2017kky,Roszkowski:2017nbc,Schumann:2019eaa}
for reviews. In particular, over the past years, direct and indirect detection experiments have reached the sensitivity necessary to probe this paradigm in many different contexts. Collider experiments also offer possibilities of tests. A number of explicit models have been already excluded, whereas many other ones could be seriously tested in the near future.

In many models WIMPs annihilate into SM fermions via 
a $t$- or $s$-channel mediator.
If this mediator is sufficiently heavy, it can be integrated out,  leading to a local effective interaction. Thus in this case the (tree level) phenomenology of the model reduces to the one that can be obtained from 
the effective field theory (EFT) 
for DM annihilation.
As is well-known too, in this case one can get a one-to-one relation between the annihilation rate fixed by the relic density constraint and direct, indirect as well as collider signals.\footnote{Actually, the criteria of having a sufficiently heavy mediator for the EFT to be valid depends on the process considered, see e.g.~\cite{Busoni:2013lha,Busoni:2014sya,Busoni:2014haa}.}

In this work we are interested in effective interactions involving charged SM fermions ($f$) and DM fermions ($\chi$), of the general form
${\cal L}\supset G\overline{\chi} {\cal O}\chi\overline{f} {\cal O}'f$ with
${\cal O}$ and ${\cal O'}$ any possible operators.
From the relic density constraint the dimensional
coupling $G$ is typically of the order of $10^{-1}G_{F}\times(m_{\chi}/100\thinspace{\rm GeV})$
where $G_{F}$ is the Fermi constant and $m_{\chi}$ is the DM mass.
If the SM fermion is a light quark and DM lies around the electroweak scale, such values of $G$  
have already been ruled out by recent direct detection experiments for  operators that lead to  spin-independent (SI) cross sections on nuclei. These are in particular the XENON-1T~\cite{Aprile:2018dbl},
LUX~\cite{Akerib:2016vxi} and PandaX-II~\cite{Cui:2017nnn} experiments which have
now put upper limits on the  SI DM-nucleon cross section
down to $\sim 10^{-46}{\rm cm}^{2}$ for $m_{\chi}$ ranging from tens
to hundreds of GeV.
However, WIMPs are not necessarily expected to  
dominantly couple to light quarks. 
For other SM fermions (e.g. $f=e,\thinspace\mu,\thinspace\tau,\thinspace c,\thinspace b,\thinspace t,\thinspace\nu_e,\thinspace\nu_\mu,\thinspace\nu_\tau$),
direct detection bounds are generally weaker. In addition, various operators may lead to spin-dependent (SD) cross sections, for which the experimental sensitivity is weaker.

An interesting possibility to improve the  direct detection sensitivity
in such cases
stems fom the fact that WIMPs might have 
electromagnetic dipole moments.  
In fact, various electromagnetic form factors (electric/magnetic dipoles, anapole, charged radius) of WIMPs have been considered in the literature~\cite{Pospelov:2000bq,Sigurdson:2004zp,Masso:2009mu,Kopp:2009et,Banks:2010eh,
Fitzpatrick:2010br,Barger:2010gv,Fortin:2011hv,
DelNobile:2012tx,Weiner:2012cb,Ho:2012bg,Gresham:2013mua,Gao:2013vfa,DelNobile:2013cva,DelNobile:2014eta,
Kopp:2014tsa,
Chu:2018qrm,Chang:2019xva,Chu:2020ysb,Ali-Haimoud:2021lka}. 
Early
studies~\cite{Fitzpatrick:2010br,Fortin:2011hv,Barger:2010gv,Banks:2010eh,Gresham:2013mua} 
considered them as a solution to resolve the discrepancy between DAMA/CoGeNT signals
and null results of other DM searches, though this has been since then well excluded. 
Collider, $\gamma$-ray, and CMB searches for dipole interacting DM have been studied in Ref.~\cite{Fortin:2011hv,DelNobile:2012tx}. 
More recently, Ref.~\cite{Kopp:2014tsa} considered leptophilic DM
and showed that its loop-induced electromagnetic dipoles led to restrictive
direct detection bounds.\footnote{Beyond the WIMP regime, there has been growing
interest in  electromagnetic dipoles of sub-GeV DM due to potential
connections with CMB/LSS observations, stellar physics, and the intensity
frontier searches---see, e.g.,  \cite{Chu:2018qrm,Chang:2019xva,Chu:2020ysb,Ali-Haimoud:2021lka}. }

In this work, instead of considering that the annihilation induced by the dipole (into SM charged particles via photon exchange) is responsible for the relic density, as in many of these works (e.g.~\cite{Masso:2009mu,Fitzpatrick:2010br,Barger:2010gv,Banks:2010eh,Fortin:2011hv}), or instead of assuming a specific model, we will instead start, as in Ref.~\cite{Kopp:2009et}, from the effective four-fermion operators. 
Once the coefficients of the effective operators are fixed
by the relic density, we can compute the 
dipoles they lead to at the one-loop level
(simply from closing the charged
fermion line, and attaching an external photon).
Actually, since dipoles are loop-level effects, the use of an effective theory to compute them is not necessarily consistent with what we would obtain in UV complete models.
 If the effective theory holds for arbitrarily high energy scales, the loop integral that leads to the dipoles would be divergent and such divergences cannot be canceled (absorbed) by any counterterms.
Unlike for the annihilation process, one thus needs to open the effective interactions. 
We consider 
two straightforward ways to generate these effective interactions at tree level, namely $s$- or $t$-channel exchange.
This leads to two general classes of models depending on whether they give a vanishing (as in the $s$-channel case) or a non-vanishing finite (as in the $t$-channel case) result.
We argue that the results obtained for non-vanishing dipoles are generic, which is illustrated by comparing these results with the ones obtained in a UV complete model.

We then study the implication of non-vanishing dipole for direct detection.
We find that the magnitude of the
non-vanishing loop-induced dipoles, typically of the order of $10^{-20}$ ${\rm ecm}$
(or $10^{-20}m_{f}/m_{\chi}$ ${\rm ecm}$), implies that DM-nucleus scattering via dipole interactions could be probed within current and future experimental sensitivities.
In particular for operators involving 
heavy quarks
or charged leptons, or when the DM-nucleus cross section is SD at
tree level, this might provide the best possibility of probing these interactions and thus possibly the origin of the DM relic density.
This stems from the fact that 
for low nuclear recoil energies
the cross section is considerably enhanced by the exchange of a massless (photon) mediator.

The paper is organized as follows. In Sec.~\ref{sec:basic}, we present a complete description of the most general four-fermion interactions of DM fermions with SM fermions, and determine the interaction strength required to produce the observed relic abundance. Given the determined interaction strength, in Sec.~\ref{sec:loop}, we compute the loop-induced electromagnetic dipoles of DM by closing charged fermion loops in the four-fermion interactions, assuming that they are induced by either $t$-channel or $s$-channel tree level exchange. There we also compare the results obtained in this way to the ones obtained from considering an explicit UV complete model.
%
In Sec.~\ref{sec:detection}, the resulting magnitude of electromagnetic dipoles is confronted  with direct detection limits obtained by investigating the recoil spectra of dipole-interacting DM.
We conclude in Sec.~\ref{sec:Conclusion} and delegate the loop calculation details to the appendix.

\section{Framework \label{sec:basic}}

\subsection{Effective interactions of DM}
\noindent
We start with  the most general four-fermion interactions of Dirac DM ($\chi$)
and SM fermions ($f$):
\begin{equation}
{\cal L}\supset G_{F}\sum_{a}\overline{\chi}\Gamma^{a}\chi\thinspace\overline{f}\Gamma^{a}(\epsilon_{a}+\tilde{\epsilon}_{a}i_{a}\gamma^{5})f\thinspace,\label{eq:m-1}
\end{equation}
where the $\Gamma^{a}$ matrices (with $a=S$, $P$, $V$, $A$, $T$) span all the 16 possible independent combinations of Dirac matrices:
\begin{equation}
\Gamma^{S}=I,\ \Gamma^{P}=i\gamma^{5},\ \Gamma^{V}=\gamma^{\mu},\ \Gamma^{A}=\gamma^{\mu}\gamma^{5},\ \Gamma^{T}=\sigma^{\mu\nu}\thinspace.\label{eq:m-2}
\end{equation}
We refer to the above five possible bi-linear products of Dirac spinors
as scalar, pseudo-scalar, vector, axial-vector, and tensor interactions.
 In Eq.~\eqref{eq:m-1}, we have normalized the interaction strength
by the Fermi constant $G_{F}$ since in the WIMP paradigm, the interactions
are typically of this magnitude. Potential deviations are absorbed
into the dimensionless constants $\epsilon_{a}$ and $\tilde{\epsilon}_{a}$.
Note that in Eq.~\eqref{eq:m-1} we have inserted an $i_{a}$ factor, which
is defined as $i_{S,P,T}=i$ and $i_{V,A}=1$, so that the various terms are hermitian,
with $\epsilon_{a}$ and $\tilde{\epsilon}_{a}$ real numbers---for
further discussions see e.g.~Refs.~\cite{Lindner:2016wff,Rodejohann:2017vup}.
For tensor interactions, one could consider adding $\gamma^{5}$ between
$\overline{\chi}$ and $\chi$ but the operator $\overline{\chi}\sigma^{\mu\nu}\gamma^{5}\chi\thinspace\overline{f}\sigma_{\mu\nu}f$
is actually identical\footnote{This can be seen as follows. In the chiral basis, one can expand it
as $\overline{\chi}\sigma^{\mu\nu}\gamma^{5}\chi\thinspace\overline{f}\sigma_{\mu\nu}f=(-\overline{\chi_{R}}\sigma^{\mu\nu}\chi_{L}+\overline{\chi_{L}}\sigma^{\mu\nu}\chi_{R})\thinspace(\overline{f_{R}}\sigma_{\mu\nu}f_{L}+\overline{f_{L}}\sigma_{\mu\nu}f_{R})$.
Since the cross terms vanish (according to Fierz transformations),
$\overline{\chi_{R}}\sigma^{\mu\nu}\chi_{L}\overline{f_{L}}\sigma_{\mu\nu}f_{R}=\overline{\chi_{L}}\sigma^{\mu\nu}\chi_{R}\overline{f_{R}}\sigma_{\mu\nu}f_{L}=0$,
the remaining terms imply $\overline{\chi}\sigma^{\mu\nu}\gamma^{5}\chi\thinspace\overline{f}\sigma_{\mu\nu}f=\overline{\chi}\sigma^{\mu\nu}\chi\thinspace\overline{f}\sigma_{\mu\nu}\gamma^{5}f$.} to $\overline{\chi}\sigma^{\mu\nu}\chi\thinspace\overline{f}\sigma_{\mu\nu}\gamma^{5}f$.
Hence Eq.~\eqref{eq:m-1} provides a complete description of all possible
Lorentz-invariant four-fermion interactions. This set of effective
operators has also been frequently used for DM searches at colliders---see e.g.~Ref.~\cite{Goodman:2010ku}. Note importantly that the $S$, $P$ and $T$ operators are not SM gauge invariant, but could be generated through electroweak symmetry breaking, see the discussion in Sec.~\ref{sub:example}.

In the SM fermion chiral basis, one can also write Eq.~\eqref{eq:m-1} as 
\begin{align}
{\cal L} & \supset G_{F}\left[\epsilon_{S}^{L}\overline{\chi}\chi\thinspace\overline{f_{R}}f_{L}+\epsilon_{P}^{L}\overline{\chi}i\gamma^{5}\chi\thinspace\overline{f_{R}}f_{L}\right.\nonumber \\
 & +\epsilon_{V}^{L}\overline{\chi}\gamma^{\mu}\chi\thinspace\overline{f_{L}}\gamma_{\mu}f_{L}+\epsilon_{A}^{L}\overline{\chi}\gamma^{\mu}\gamma^{5}\chi\thinspace\overline{f_{L}}\gamma_{\mu}f_{L}\nonumber \\
 & +\left.\epsilon_{T}^{L}\overline{\chi}\sigma^{\mu\nu}\chi\thinspace\overline{f_{R}}\sigma_{\mu\nu}f_{L}+(L\leftrightarrow R)\right],\label{eq:m-8}
\end{align}
where $f_{L,R}\equiv P_{L,R}f$, $P_{L,R}\equiv(1\mp\gamma^{5})/2$,
$\epsilon_{a}^{L}$ and $\epsilon_{a}^{R}$ are linear combinations
of $\epsilon_{a}$ and $\tilde{\epsilon}_{a}$. Given the chiral structure of the SM, and the fact that most results are symmetric under $L\leftrightarrow R$, in this work we will adopt
the chiral basis. Note that while $\epsilon_{a}$ and $\tilde{\epsilon}_{a}$
in Eq.~\eqref{eq:m-1} are real and independent of each other, $\epsilon_{a}^{L}$
and $\epsilon_{a}^{R}$ in the chiral basis are either complex conjugate
of each other ($\epsilon_{S}^{R}=\epsilon_{S}^{L*}$, $\epsilon_{P}^{R}=\epsilon_{P}^{L*}$,
$\epsilon_{T}^{R}=\epsilon_{T}^{L*}$), or real and independent ($\epsilon_{V}^{R}={\rm Re}\left[\epsilon_{V}^{R}\right]$,
$\epsilon_{V}^{L}={\rm Re}\left[\epsilon_{V}^{R}\right]$, $\epsilon_{A}^{R}={\rm Re}\left[\epsilon_{A}^{R}\right]$,
$\epsilon_{A}^{L}={\rm Re}\left[\epsilon_{A}^{R}\right]$). Hence
the full  set of $\epsilon$'s in the chiral basis still contains
10 real independent parameters.

\subsection{DM relic abundance}
\noindent
The relic abundance of $\chi$ via the standard freeze-out mechanism
is approximately given by (see e.g. \cite{Plehn:2017fdg})
\begin{equation}
\Omega_{\chi}h^{2}\simeq0.12\frac{x_{{\rm f.o}}}{23}\frac{\sqrt{g_{\star}}}{10}\frac{1.7\times10^{-9}{\rm GeV}^{-2}}{\langle\sigma v\rangle},\label{eq:m-6-1}
\end{equation}
where $x_{{\rm f.o}}\equiv T_{{\rm f.o}}/m_{\chi}$ is the ratio of
the freeze-out temperature $T_{{\rm f.o}}$ to the WIMP mass $m_{\chi}$;
$g_{\star}$ is the effective number of relativistic degrees of freedom in the thermal bath at freeze-out;
 and $\langle\sigma v\rangle$ is defined as~\cite{Kolb,Gondolo:1990dk}
\begin{equation}
\langle\sigma v\rangle\equiv n_{{\rm EQ}}^{-2}\int|{\cal M}|^{2}d\Pi_{1}d\Pi_{2}d\Pi_{3}d\Pi_{4}(2\pi)^{4}\delta^{4}f_{1}f_{2}\thinspace,\label{eq:m-3}
\end{equation}
\vspace{-0.5cm}
\begin{equation}
n_{{\rm EQ}}\equiv\int2E_{1}d\Pi_{1}f_{1}\thinspace,\ \ d\Pi_{i}\equiv\frac{g_{i}d^{3}\mathbf{p}_{i}}{(2\pi)^{3}2E_{i}}\thinspace.\label{eq:m-5}
\end{equation}
Here subscripts $1$, $2$, $\cdots$, and 4 denote quantities of
the first, second, $\cdots$, and the fourth particles in $\chi+\overline{\chi}\rightarrow f+\overline{f}$;
$\delta^{4}$ is short for $\delta^{4}(p_{1}+p_{2}-p_{3}-p_{4})$;
and  $f_{1}$ ($f_{2}$) is the thermal distribution function of $\chi$
($\overline{\chi}$). The squared amplitudes $|{\cal M}|^{2}$ has been evaluated and summarized in Tab.~\ref{tab:t}.

For $P$, $V$, $T$ interactions, the annihilation amplitudes are of the s-wave type and consequently are
nearly constant in the non-relativistic regime. 
In this case, we can neglect the velocity dependence and
reduce Eq.~\eqref{eq:m-3} to
\begin{align}
\langle\sigma v\rangle & \equiv\frac{1}{4m_{\chi}^{2}}\int|{\cal M}|^{2}d\Pi_{3}d\Pi_{4}(2\pi)^{4}\delta^{4}=\frac{|{\cal M}|^{2}}{32\pi m_{\chi}^{2}}.\label{eq:m-4}
\end{align}

For $S$ and $A$ interactions, we have $|{\cal M}|^{2}\propto v^{2}$
($p$-wave annihilation) and hence the integration is somewhat more complicated.
Assuming Maxwell-Boltzmann distributions for $f_{1}$ and $f_{2}$,
Eq.~\eqref{eq:m-3} can be reduced to~\cite{Gondolo:1990dk}
\begin{equation}
\langle\sigma v\rangle\equiv\frac{1}{8m_{\chi}^{4}TK_{2}^{2}\left(m_{\chi}/T\right)}\int_{4m_{\chi}^{2}}^{\infty}\sigma\sqrt{s}\left(s-4m_{\chi}^{2}\right)K_{1}\left(\sqrt{s}/T\right)ds.\label{eq:m-33}
\end{equation}
Here $K_{1}$ and $K_{2}$ are $K$-type Bessel function of orders
1 and 2, $s=(p_{1}+p_{2})^{2}=4m_{\chi}^{2}+m_{\chi}^{2}v^{2}$, and
$\sigma$ is the total annihilation cross section~\cite{Zyla:2020zbs}:
\begin{equation}
\sigma=\int\frac{|{\cal M}|^{2}}{16\pi s\left(s-4m_{\chi}^{2}\right)}dq^{2},\label{eq:m-34}
\end{equation}
where $q^{2}=(p_{3}-p_{1})^{2}\approx-m_{\chi}^{2}(1-v\cos\theta)$,
with $\theta$ the angle between $\mathbf{p}_{1}$ and $\mathbf{p}_{3}$.
Integrating $q^{2}$  from $-m_{\chi}^{2}(1+v)$ to $-m_{\chi}^{2}(1-v)$,
 we obtain results for $\sigma$ that are given in Tab.~\ref{tab:t}. 

Plugging the results for $\sigma$ into Eq.~\eqref{eq:m-33} with
$v\rightarrow m_{\chi}^{-1}\sqrt{s-4m_{\chi}^{2}}$, we can integrate
Eq.~\eqref{eq:m-33} analytically by noticing that for any value of $p>-1$,
\begin{equation}
\int_{4m^{2}}^{\infty}\left(s-4m^{2}\right)^{p}K_{1}\left(\sqrt{s}/T\right)\frac{1}{\sqrt{s}}ds=2^{1+2p}T(mT)^{p}K_{p}\left(2m/T\right)\Gamma(1+p),\label{eq:m-35}
\end{equation}
where $\Gamma$ is the Euler gamma function. The results for $\langle\sigma v\rangle$
are then expanded in $T/m$ and summarized in Tab.~\ref{tab:t}.

Using the results for $\langle\sigma v\rangle$ with $T_{\rm f.o}\simeq m_{\chi}/23$
(the typical freeze-out temperature) in Eq.~\eqref{eq:m-6-1}, we
obtain 
\begin{equation}
\Omega_{\chi}h^{2}\simeq0.12\left(\frac{100\ \text{GeV}}{m_{\chi}}\right)^{2}\left|\frac{\epsilon_{a}^{\star}}{\epsilon_{a}^{L,R}}\right|^{2},\label{eq:m-7}
\end{equation}
where $\epsilon_{a}^{\star}$ denotes benchmark values: $\epsilon_{S}^{\star}=0.49$,
$\epsilon_{P}^{\star}=0.13$, $\epsilon_{V}^{\star}=0.089$, $\epsilon_{A}^{\star}=0.43$,
and $\epsilon_{T}^{\star}=0.063$. 
Note that  $\epsilon_{a}^{\star}$ for $a=S$ or $A$ is generally larger than for other cases because the cross section is velocity suppressed, which implies that they would freeze out at higher temperatures for the same coupling strength. Hence to reach the same relic abundance (i.e. same freeze out temperature), the coupling needs to be larger.  

\newcommand{\vvvv}{ \rule[-2ex]{0pt}{6ex} }

\begin{table*}
\caption{\label{tab:t} Annihilation amplitudes ($|{\cal M}|^{2}$), cross
sections ($\sigma$), thermally averaged cross sections ($\langle\sigma v\rangle$),
and benchmark values of $\epsilon_{a}^{\star}$ used in Eq.~\eqref{eq:m-7}  for the five types
of effective interactions. We neglected the mass of the final states
and assume that the annihilating DM particles are non-relativistic with $v$
being their relative velocity. The results have been expanded in $v$
and only leading-order terms are retained. Results for $\epsilon_{a}^{R}$
and $\epsilon_{a}^{L}$ are identical.}

\centering
\begin{tabular}{ccccccc}
\hline \hline

 & $S$ & $P$ & $V$ & $A$ & $T$ & \tabularnewline
\hline 
\vvvv $|{\cal M}|^{2}/\left|G_{F}m_{\chi}^{2}\epsilon_{a}^{L,R}\right|^{2}$
 & $2v^{2}$ & $8$ & $16$ & $2v^{2}(1+\cos^2\theta)$ & $32$ & \tabularnewline
\vvvv $\sigma/\left|G_{F}m_{\chi}^{2}\epsilon_{a}^{L,R}\right|^{2}$ & $\frac{v}{4\pi s}$ & $\frac{1}{\pi sv}$ & $\frac{2}{\pi sv}$ & $\frac{v}{3\pi s}$ & $\frac{4}{\pi s v}$ & \tabularnewline
\vvvv $\langle\sigma v\rangle/\left|G_{F}m_{\chi}^{2}\epsilon_{a}^{L,R}\right|^{2}$ & $\frac{3}{8\pi}Tm_{\chi}^{-3}$ & $\frac{1}{4\pi}m_{\chi}^{-2}$ & $\frac{1}{2\pi}m_{\chi}^{-2}$ & $\frac{1}{2\pi}Tm_{\chi}^{-3}$ & $\frac{1}{\pi}m_{\chi}^{-2}$ & \tabularnewline
$\epsilon_{a}^{\star}$ & $0.49$ & $0.13$ & $0.089$ & $0.43$ & $0.063$ & \tabularnewline
\hline \hline
\end{tabular}

\end{table*}

\section{Loop-induced electromagnetic interactions\label{sec:loop}}

\subsection{Closing the loop}

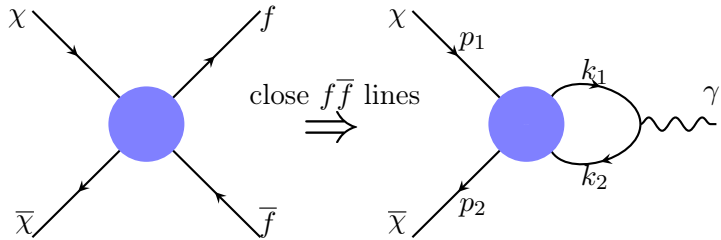
\begin{figure}
\centering

\usetikzlibrary{decorations.pathmorphing,decorations.markings}

\begin{tikzpicture}

\node [text=black,rotate=0] at (-4.2,1.9) {$\chi$};
\node [text=black,rotate=0] at (-0.9,1.9) {$f$};
\node [text=black,rotate=0] at (-4.1,-0.8) {$\overline{\chi}$};
\node [text=black,rotate=0] at (-0.9,-0.8) {$\overline{f}$};

\begin{scope}[thick,decoration={
    markings,
    mark=at position 0.5 with {\arrow{stealth}}}
    ]

	\draw[postaction={decorate}] (-2.8,0.2)--(-4,-1);
    \draw[postaction={decorate}] (-2.2,0.8)--(-1,2);
    \draw[postaction={decorate}] (-4,2)--(-2.8,0.8);
    \draw[postaction={decorate}] (-1,-1)--(-2.2,0.2);
   
\end{scope}
\fill[blue!50] (-2.5,0.5)circle (0.5);

\begin{scope}[thick,decoration={
    markings,
    mark=at position 0.5 with {\arrow{stealth}}}
    ]

	\draw[postaction={decorate}] (2.2,0.2)--(1,-1);
    \draw[postaction={decorate}] (1,2)--(2.2,0.8);
     \draw [postaction={decorate}] (2.8,0.8) to[out=70,in=90] (4.0,0.5) ;
      \draw [postaction={decorate}] (4.0,0.5) to[in=-70,out=-90]  (2.8,0.2);
	\draw [decorate,decoration=snake](4,0.5)--(5,0.5);
\end{scope}

\fill[blue!50] (2.5,0.5)circle (0.5);
\node [text=blue!40,rotate=0,scale=0.05] at (2.5,0.5) {19931126};
\node [text=black,rotate=0] at (4.9361,0.8747) {$\gamma$};

\node [text=black,rotate=0] at (0.8,1.9) {$\chi$};
\node [text=black,rotate=0] at (0.8,-0.8) {$\overline{\chi}$};
\node [text=black,rotate=0] at (1.8,1.6) {$p_1$};
\node [text=black,rotate=0] at (1.8,-0.6) {$p_2$};

\node [text=black,rotate=0] at (3.4,1.2) {$k_1$};
\node [text=black,rotate=0] at (3.4,-0.2) {$k_2$};

\node  [text=black,rotate=0] at (-0.0195,0.8988) {close $f\overline{f}$ lines};
\node  [text=black,rotate=0,scale=2] at (-0.0677,0.4105) {$\Rightarrow$ };
\end{tikzpicture}

\caption{Loop-induced electromagnetic interactions. In the presence of effective
$\overline{\chi}$-$\overline{\chi}$-$\overline{f}$-$f$ interactions
where $\chi$ is a DM fermion and $f$ is a SM fermion, the $f$ and
$\overline{f}$ lines can be closed to form a loop diagram, which
after attaching a photon line can generate electric and/or magnetic
dipoles of DM. \label{fig:close-f}}
\end{figure}

\noindent In the presence of any of the $\overline{\chi}$-$\chi$-$\overline{f}$-$f$
interactions formulated in Eq.~\eqref{eq:m-1} or Eq.~\eqref{eq:m-8},
we can close the fermion lines of $f$ and attach a photon external
line, as illustrated in Fig.~\ref{fig:close-f}. This generally leads
to loop-induced electromagnetic interactions of $\chi$. 

Closing the loop in this way one gets amplitudes which take the general form
\begin{align}
i{\cal M}_{{\rm loop}} & =i\overline{u_{2}}\Gamma^{a}u_{1}\int\frac{d^{4}k}{(2\pi)^{4}}{\rm tr}\left[\frac{1}{\slashed{k}_{2}-m_{f}}i eQ_{f}\gamma^{\mu}\frac{1}{\slashed{k}_{1}-m_{f}}\tilde{\Gamma}^{a}\right]\varepsilon_{\mu}
G_{X}(p_{1},p_{2},k_{1},k_{2})\nonumber \\
 & \equiv i\overline{u_{2}}{\cal F}^{\mu}u_{1}\varepsilon_{\mu},\label{eq:m-9}
\end{align}
where $\overline{u_{2}}$, $u_{1}$, $\varepsilon_{\mu}$ represent
the three external lines; the momenta $p$'s and $k$'s  have been
defined in Fig.~\ref{fig:close-f} with  $k=p_{1}-k_{1}=p_{2}-k_{2}$;
$m_{f}$ and $Q_{f}$ are the mass and the electric charge of $f$;
$\tilde{\Gamma}^{a}\equiv\Gamma^{a}P_{L,R}$; $G_{X}$ is the coefficient of the effective operator considered.
The most general form of the ${\cal F}^{\mu}$ vertex function
that respects Lorentz and electromagnetic gauge invariance can be
decomposed as a combination of four terms, each one with its own form factor (see e.g.~\cite{Nowakowski:2004cv,Giunti:2014ixa}):
\begin{equation}
{\cal F}^{\mu}={\cal F}_{Q}(q^{2})\gamma^{\mu}+{\cal F}_{M}(q^{2})i\sigma^{\mu\nu}q_{\nu}+{\cal F}_{E}(q^{2})\sigma^{\mu\nu}\gamma^{5}q_{\nu}+{\cal F}_{A}(q^{2})(q^{2}\gamma_{\mu}-q_{\mu}\slashed{q})\gamma^{5}.\label{eq:m-10}
\end{equation}
Here $q\equiv p_{1}-p_{2}$ and in the limit of $q^{2}\rightarrow0$,
the four form factors ${\cal F}_{Q}(0)$, ${\cal F}_{M}(0)$, ${\cal F}_{E}(0)$,
and ${\cal F}_{A}(0)$ are the electric charge, magnetic dipole, electric
dipole and anapole of $\chi$, respectively. For simplicity, we denote
\begin{equation}
d_{M}\equiv{\cal F}_{M}(0),\ \ d_{E}\equiv{\cal F}_{E}(0).\label{eq:m-15}
\end{equation}
In this work, we do not consider the electric charge and anapole of
$\chi$ because the former remains zero at loop levels if DM is electrically neutral
at tree level and the latter causes suppressed signals in DM direct
detection. This suppression can be seen from the form of the ${\cal F}_{A}$
term of Eq.~\eqref{eq:m-10}, which in the low-$q^{2}$ regime is proportional to 
$q^{2}$. This ${\cal O}(q^{2})$ coefficient  will be canceled
by the photon propagator which is proportional to ${\cal O}(q^{-2})$.
Indeed, Ref.~\cite{Kopp:2014tsa} has shown that the effect of anapole
in direct detection is nearly equivalent to that of contact interactions.
Thus, unlike with dipoles, the direct detection does not profit from the several orders of magnitude enhancement related to the $1/q^2$ behavior of the amplitude, see below.
Neutral $\chi$ might possess a non-vanishing charge radius defined
as $d{\cal F}_{Q}(q^{2})/dq^{2}|_{q^{2}\rightarrow0}$. Its effect
in direct detection is also suppressed for the same reason.\footnote{For a scalar DM candidate the coupling to the photon that could be loop induced from an effective operator (i.e.~$\phi_{DM}^\dagger \phi_{DM} \bar{f} f$ or $\phi_{DM}^\dagger \phi_{DM} \bar{f} \gamma_5 f$), coupling it in pairs to a pair of charged fermions, would lead to suppressed direct detection in a similar way, as it would induce only a charged radius. For a Majorana DM candidate, as is well known, dipole interactions identically vanish, and an anapole leads to suppressed direct detection in a similar way than for a Dirac fermion.
} 

% ... 


\newcommand{\vvv}{ \rule[-3ex]{0pt}{6ex} }

\begin{table*}
\caption{\label{tab:dipole} Electromagnetic dipoles generated by the loop
diagram in Fig.~\ref{fig:close-f}. Here the ``$t$'' and ``$s$'' indices refer to
the results obtained considering the corresponding channel 
in Eq.~\eqref{eq:m-14}.
$C^{(t)}$ and $C^{(s)}$ are given in Eq.~\eqref{eq:Cts}.
 }

    \centering
    \resizebox{\textwidth}{!}{
    \begin{tabular}{ccccccc}
    \hline \hline
     & $S$ & $P$ & $V$ & $A$ & $T$ & \tabularnewline
    \hline 
    \vvv $d_{M}^{(t)}/\left(\frac{eQ_{f}G_{F}}{16\pi^{2}}\right)$  & $(\epsilon_{S}^{L}+\epsilon_{S}^{R})m_{f}$ & 0 & $\frac{1}{3}(\epsilon_{V}^{L}+\epsilon_{V}^{R})m_{\chi}$ & $-(\epsilon_{A}^{L}-\epsilon_{A}^{R})m_{\chi}$ & $4(\epsilon_{T}^{L}+\epsilon_{T}^{R})m_{f}C^{(t)}$ & \tabularnewline
    \vvv  $d_{E}^{(t)}/\left(\frac{eQ_{f}G_{F}}{16\pi^{2}}\right)$ & 0 & $(\epsilon_{P}^{L}+\epsilon_{P}^{R})m_{f}$ & 0 & 0 & $4i(\epsilon_{T}^{L}-\epsilon_{T}^{R})m_{f}C^{(t)}$ & \tabularnewline
    \vvv  $d_{M}^{(s)}/\left(\frac{eQ_{f}G_{F}}{16\pi^{2}}\right)$ & 0 & 0 & 0 & 0 & $4(\epsilon_{T}^{L}+\epsilon_{T}^{R})m_{f}C^{(s)}$ & \tabularnewline
    \vvv  $d_{E}^{(s)}/\left(\frac{eQ_{f}G_{F}}{16\pi^{2}}\right)$ & 0 & 0 & 0 & 0 & $4i(\epsilon_{T}^{L}-\epsilon_{T}^{R})m_{f}C^{(s)}$ & \tabularnewline
    \hline \hline
    \end{tabular}
    }



\end{table*}

Since they involve a loop where momenta runs from 0 to infinity, the use of the effective theory to compute these dipoles does not necessarily lead to consistent results. 
Related to that, two explicit UV complete theories leading to the same operators at low energy does not necessarily lead to the same dipoles. Explicit calculations shows that indeed the calculation of the dipoles from Fig.~\ref{fig:close-f}, i.e.~with $G_X$ a constant, is not consistent, since it leads to loop integral divergent results. Thus one must open the four-fermion interactions.
Here we will open them along the two simplest possible ways, from the exchange of a $t$-channel or $s$-channel heavy mediator giving momentum-dependent $G_X$ functions
\begin{equation}
G_{X}^{(s)}=\frac{y^{2}}{s-m_{\text{med}}^{2}},\ \ G_{X}^{(t)}=\frac{y^{2}}{t-m_{\text{med}}^{2}},\label{eq:m-14}
\end{equation}
where $m_{\text{med}}$ and $y$ are the mediator mass and coupling,
$s=(p_{1}-p_{2})^{2}$ and $t=(p_{1}-k_{1})^{2}$.\footnote{Note that when
the left diagram is interpreted as DM annihilation, we flip the direction
of $p_{2}$ and obtain $s=(p_{1}+p_{2})^{2}$ which is the conventional
definition of $s$ as the Mandelstam variable.} There could be more
complex scenarios for the internal structure of the effective vertex,
where e.g. $G_{X}$ is generated by a box diagram, for which the current framework does not apply since it typically requires two-loop calculations (which are beyond the scope
of this work).

Substituting Eq.~\eqref{eq:m-14} in Eq.~\eqref{eq:m-9} and performing
the loop integration (see Appendix~\ref{sec:loop-cal}), 
we obtain the results in Tab.~\ref{tab:dipole} where $C^{(t)}$ and $C^{(s)}$ in the last column are defined as 
\begin{equation}
    C^{(t)}\equiv 1+\log(m_{f}^{2}/m_{{\rm med}}^{2}),\ \ C^{(s)}\equiv\frac{1}{\varepsilon}+\log(\mu^{2}/m_{f}^{2}).\label{eq:Cts}
\end{equation}
Here $ C^{(t)}$ is finite but $C^{(s)}$ contains a UV divergence, with $\mu$ and $\varepsilon$ defined by 
the dimensional regularization $d^{4}k/(2\pi)^{4}\rightarrow\mu^{2\varepsilon}d^{4-2\varepsilon}k\thinspace/(2\pi)^{4-2\varepsilon}$.\footnote{Note however that the tensor operator cannot result from a simple tree level $s$-channel exchange, but must be induced e.g.~from a loop diagram coupling the $s$-channel mediator to the pair of DM particles and from another loop diagram coupling the $s$-channel mediator to the pair of SM fermions. This case is thus of limited interest. Tensor interactions are generated in an easier way from tree level $t$-channel (through Fierz transformation) or one loop box diagrams.
Thus, even if this divergence, that we get only for this $s$-channel tensor case, means that the result is inconsistent  (and that in UV complete models this dipole interaction necessarily never comes without other interactions), we will not elaborate more on this problem. At a very rough level one can expect constraints on this case similar to the ones obtained below for the t-channel $T$ case.} 
The results are obtained assuming the heavy mediator limit: $m_{\text{med}}\gg m_{\chi,f}$. 
It is noteworthy that the loop-induced dipoles for the $S$, $P$ and $T$ cases are proportional to $m_{f}$ while for the other two cases they are proportional to $m_{\chi}$. This is due to the well-known chirality-flipping nature of $S$, $P$, $T$ interactions---see
discussions in Ref.~\cite{Xu:2019dxe}.

For the $t$-channel case a magnetic or an electric 
dipole is always induced (even if never both), depending on the operator considered. In all cases this allows non-suppressed direct detection signals as we will see below. Baring cancellations this implies that any UV complete model generating any one of these operators through a $t$-channel transition can be efficiently probed via direct detection, see below. This is presumably also the case for models where the effective interactions would be induced at loop level, such as through box diagrams (but we will not explicitly check this statement here).
Note that for this $t$-channel case all the results are obtained finite as it should obviously be.
In the $s$-channel case instead no dipoles at all are obtained for the $S$, $P$, $V$ and $A$ cases, as a result of the fact that in this case the loop is a self-energy which cannot give rise to a $\sigma_{\mu\nu}$. For the tensor case one can get a dipole as the effective operator already contain a $\sigma_{\mu\nu}$ to start with. 

In summary, we get two general classes of scenarios, the one leading for simple reasons to vanisihing dipoles and the ones leading to non-vanishing dipoles. For the second class we have 
\begin{equation}
d_{M,E}=\frac{eQ_{f}G_{F}}{16\pi^{2}}\times{\cal O}(\epsilon_{a})\times\begin{cases}
m_{\chi} & \text{for }a=V,\ A\\
m_{f} & \text{for }a=S,\ P,\ T
\end{cases},\label{eq:m-36}
\end{equation}
where the ${\cal O}(\epsilon_{a})$ part has been specified in Tab.~\ref{tab:dipole}.

\subsection{A UV complete example\label{sub:example}}
\noindent 

The results obtained above for the dipoles by replacing in Fig.~\ref{fig:close-f} the four fermion interaction by a $t$-channel propagator have no reasons to give exactly what we would get in a UV complete model leading to these operators through a $t$-channel heavy mediator exchange.
Since DM is neutral and the SM fermions are charged, the $t$-channel heavy mediator has necessarily a non-vanishing electric charge. Thus in a UV complete model there are necessarily extra diagrams that may modify the dipoles, simply attaching the photon to the heavy mediator rather than to the SM charged fermion. However, for a mediator much heavier than the other particles we do not expect in general that these extra diagrams could induce any large destructive interference for the dipole induced (given in particular the chiral structure of the SM), or even largely change the results.
To illustrate this, we consider a simple UV complete model leading to the 4-fermion interactions through $t$-channel exchange.
Consider a charged
scalar $\phi^{\pm}$ that couples to $f_R$ and $\chi$: 
\begin{equation}
{\cal L}\supset y\overline{\chi}f_R\phi^{+}+{\rm h.c.}\label{eq:x-8}
\end{equation}
Assuming the scalar boson mass $m_{\phi}$ is heavy, by integrating
out $\phi^{\pm}$, we obtain the effective interaction
\begin{equation}
{\cal L}_{{\rm eff}}=-G_{X}\overline{\chi}P_R f\,\overline{f}P_L\chi,\label{eq:x-9}
\end{equation}
where $G_{X}=yy^{*}/m_{\phi}^{2}$. One can reformulate it to the form in Eq.~\eqref{eq:m-1}
via Fierz transformation:\footnote{See, e.g., Ref.~\cite{Giunti}, page 65.}
\begin{align}
    {\cal L}_{{\rm eff}}= &  \frac{1}{4}G_{X}\overline{\chi}\gamma^{\mu}\chi\overline{f_R}\gamma_{\mu}f_R-\frac{1}{4}G_{X}\overline{\chi}\gamma^{\mu}\gamma^{5}\chi\overline{f_R}\gamma_{\mu}f_R.\label{eq:x-10}
\end{align}
Eq.~\eqref{eq:x-10} contains two types ($V$ and $A$)
of effective interactions, with the following $\epsilon$'s:\footnote{It is not surprising that we get a combination of the SM gauge group invariant $V$ and $A$ operators since the original interactions of Eqs.~\eqref{eq:x-8} and \eqref{eq:x-9} are gauge invariant. Operators of the $S$, $P$ or $T$ type can be generated in UV complete models, for instance from inducing a gauge invariant dimension 7 operator involving an extra Higgs doublet and electroweak symmetry breaking. In this case the extra interactions involving the Higgs doublet components rather than the Higgs boson vev do not induce an extra contribution to the dipole at same one loop order (but could be relevant for DM annihilation for multi-TeV DM, i.e.~for $m_{DM}\gg v_{EW}$). Such operators could also be generated if for instance, on top of the interaction of Eq.~\eqref{eq:x-8}, there exists a similar interaction involving a SM left-handed doublet and a scalar doublet rather than $f_R$ and $\phi^+$, and if the charged component of this scalar doublet mixes with $\phi^+$ via electroweak symmetry breaking.} 
\begin{align}
    &\epsilon_{V}^{L}=0,\ \epsilon_{V}^{R}=\frac{1}{4}G_{X}/G_{F},\nonumber \\
    & \epsilon_{A}^{L}=0,\ \epsilon_{A}^{R}=-\frac{1}{4}G_{X}/G_{F}.\label{eq:x-13}
   \end{align}
According to Tab.~\ref{tab:dipole}, by summing up the contributions of the above $\epsilon$'s, i.e.~from Eq.~\eqref{eq:x-10} and Tab.~\ref{tab:dipole}, 
we obtain the magnetic dipole in the EFT approach:
\begin{equation}
    d^{\rm EFT}_{M}=-\frac{1}{6}m_{\chi} \cdot \frac{eG_{X}Q_{f}}{16\pi^{2}}.\label{eq:x-14}
\end{equation}

One can also compute exactly the dipoles directly from the model Lagrangian.
As already mentioned above, there are two diagrams that couple $\chi$-$\overline{\chi}$ to the photon, one with the photon coupled to $f$ and the other one with the photon coupled to $\phi^{\pm}$. The resulting magnetic dipole from the two diagrams reads (see Appendix~\ref{sec:loop-cal} for details on this calculation):
\begin{equation}
    d^{\rm UV}_{M}=-\frac{1}{4}m_{\chi}\cdot \frac{eG_{X}Q_{f}}{16\pi^{2}}.\label{eq:dMUV}
\end{equation}
Note that a purely left-handed interaction involving a SM fermion doublet and a scalar doublet instead of $f_R$ and $\phi^+$ in Eq.~\eqref{eq:x-8} gives the same dipoles as the purely right-handed case of Eq.~\eqref{eq:x-8}.


By comparing Eq.~\eqref{eq:dMUV} to Eq.~\eqref{eq:x-14}, we see that the 
magnetic dipole computed in the UV theory in actually  50\% higher than the EFT result.
Therefore, when using the results of Tab.~\ref{tab:dipole}, one should keep in mind that the results may be changed  due to new contributions in complete theories. Nevertheless, the EFT driven results of Tab.~\ref{tab:dipole} provide correct estimates of the order of magnitude of the dipoles.  In other words, one cannot exclude that specific UV models would give quite different results between both approaches but this explicit example shows that in simple frameworks this is not the case.

\subsection{Expected magnitude of electromagnetic dipoles}

\noindent The relic abundance constraint, $\Omega_{\chi}h^{2}\simeq0.12$, requires that the $\epsilon$ coefficients 
of the effective operators are
typically of the order of a few (or tens of ) percent---see Tab.~\ref{tab:t}
and Eq.~\eqref{eq:m-7}. By requiring that $\Omega_{\chi}h^{2}\simeq0.12$
is correctly produced, according to Eq.~\eqref{eq:m-7}, we replace
$\epsilon_{a}^{L,R}$ in Tab.~\ref{tab:dipole} with $100\ \text{GeV}\cdot m_{\chi}^{-1}\epsilon_{a}^{\star}$
and obtain
\begin{equation}
|d_{M}^{(t)}|\approx\begin{cases}
4.3\times10^{-21}|Q_{f}|\thinspace{\rm ecm} & \text{for }a=V\\
6.2\times10^{-20}|Q_{f}|\thinspace{\rm ecm} & \text{for }a=A
\end{cases},\label{eq:m-37}
\end{equation}
where ${\rm ecm}\equiv e\times{\rm cm}\approx15350.3\ {\rm eV}^{-1}$
is a commonly used unit for electromagnetic dipoles. The result is
almost independent of $m_{\chi}$ and $m_{f}$. It only depends on
the electric charge of the SM fermion involved, $Q_{f}$, which can
be $2/3$ (for $f=u$, $c$, $t$), $-1/3$ (for $f=d$, $s$, $b$),
or $-1$ (for $f=e$, $\mu$, $\tau$).

For $S$, $P$, and $T$ interactions, the dipoles depend on $m_{f}$.
Since in this work we require that $\chi\overline{\chi}\rightarrow f\overline{f}$
is responsible for the relic abundance, we concentrate on cases with
$m_{\chi}\gtrsim m_{f}$. With this assumption,  one can still
apply $\epsilon_{a}^{L/R}=100\ \text{GeV}\cdot m_{\chi}^{-1}\epsilon_{a}^{\star}$
to the remaining dipoles:
\begin{align}
|d_{M}^{(t)}| & \approx7.2\times10^{-20}|Q_{f}|\frac{m_{f}}{m_{\chi}}\thinspace{\rm ecm},\ \ \text{for }a=S\thinspace,\label{eq:m-38}\\
|d_{E}^{(t)}| & \approx1.8\times10^{-20}|Q_{f}|\frac{m_{f}}{m_{\chi}}\thinspace{\rm ecm},\ \ \text{for }a=P\thinspace,\label{eq:m-39}\\
|d^{(t)}_{E,M}|& \approx3.7\times10^{-20}|Q_{f}|\frac{m_{f}}{m_{\chi}}\thinspace{\rm ecm},\ \ \text{for }a=T\thinspace.\label{eq:m-40}
\end{align}

\section{Electromagnetic dipoles in direct detection\label{sec:detection}}

\noindent In direct detection experiments, the differential event
rate of DM-nucleus scattering\footnote{
    In Ref.~\cite{Kopp:2009et}, it was pointed out for leptophilic DM that
    DM-nucleus scattering largely dominates 
    over DM-electron scattering (for DM masses beyond GeV) and this is expected here too for the same reasons.} can be evaluated via (see e.g. \cite{Lin:2019uvt,Zyla:2020zbs}):
\begin{equation}
\frac{dR}{dE_{r}}=N_{T}n_{\chi}\epsilon(E_{r})\int\frac{d\sigma}{dE_{r}}vf_{\oplus}(\mathbf{v})\Theta(v-v_{\min})d^{3}\mathbf{v}\thinspace.\label{eq:m-20}
\end{equation}
Here $E_{r}$ is the nuclear recoil energy; $N_{T}$ is the total
number of target nuclei; $n_{\chi}$ is the local DM number density;
$\epsilon(E_{r})$ is the detection efficiency; $\frac{d\sigma}{dE_{r}}$
is the differential cross section; $f_{\oplus}(\mathbf{v})$ is the
DM velocity distribution in the Earth frame; $v_{\min}$ is the minimal
velocity to generate a given $E_{r}$,
\begin{equation}
v_{{\rm min}}=\sqrt{\frac{m_{N}E_{r}}{2\mu_{\chi N}^{2}}}\thinspace,\label{eq:m-19}
\end{equation}
where $m_{N}$ is the nucleus mass and $\mu_{\chi N}\equiv m_{\chi}m_{N}/(m_{\chi}+m_{N})$
is the DM-nucleus reduced mass. For the local DM density we take $n_{\chi}=\rho_{\chi}/m_{\chi}$
and $\rho_{\chi}=0.4\ {\rm GeV}/\text{cm}^{3}$~\cite{Read:2014qva}. 

The DM velocity distribution $f_{\oplus}(\mathbf{v})$ is often parametrized
by a truncated Maxwellian distribution in the frame of the Galaxy
and then boosted to the Earth frame. The specific form reads
\begin{equation}
f_{\oplus}(\mathbf{v})=\frac{1}{N_{f}}\exp\left[-\frac{\tilde{v}^{2}}{v_{0}^{2}}\right]\Theta(v_{{\rm esc}}-\tilde{v})\thinspace,\label{eq:m-21}
\end{equation}
where $\tilde{v}=|\mathbf{v}+\mathbf{v}_{\oplus}|$ and $|\mathbf{v}_{\oplus}|\approx240\ \text{km}/\text{s}$
is the velocity of the Earth with respect to the Galaxy; $v_{{\rm esc}}\approx550\ \text{km}/\text{s}$
is the escape velocity of the Galaxy; $v_{0}=220\ \text{km}/\text{s}$
is the mean velocity of the Maxwellian distribution. The $N_{f}$
factor normalizes $f_{\oplus}$ so that $\int f_{\oplus}(\mathbf{v})d^{3}\mathbf{v}=1$:
\begin{equation}
N_{f}=\pi^{3/2}v_{0}^{3}\left[{\rm erf}\left(\frac{v_{{\rm esc}}}{v_{0}}\right)-\frac{2}{\sqrt{\pi}}\frac{v_{{\rm esc}}}{v_{0}}\exp\left(-\frac{v_{{\rm esc}}^{2}}{v_{0}^{2}}\right)\right]\thinspace.\label{eq:m-22}
\end{equation}

The differential cross sections for DM-nucleus scattering via dipoles
read~\cite{Banks:2010eh,DelNobile:2012tx}:
\begin{align}
\frac{d\sigma_{M}}{dE_{r}} & \approx d_{M}^{2}\frac{\alpha Z^{2}\text{ }F_{E}^{2}}{E_{r}}\left[1-\frac{E_{r}}{2m_{N}v^{2}}\left(1+2\frac{m_{N}}{m_{\chi}}\right)\right]+d_{M}^{2}\frac{\alpha GF_{M}^{2}}{2m_{N}v^{2}}\thinspace,\label{eq:m-23}\\
\frac{d\sigma_{E}}{dE_{r}} & \approx d_{E}^{2}\frac{\alpha Z^{2}}{E_{r}v^{2}}F_{E}^{2}\thinspace.\label{eq:m-24}
\end{align}
Note the $1/E_r$ dependence of these differential cross sections, stemming from the propagator of the (massless) photon. As compared to a standard WIMP case, where the particle exchanged with the nucleon has typically an electroweak scale mass, this will largely boost the number of events in direct detection events, since the recoil energy considered in these experiments is typically of order $5$-$50$~keV (see below). This explains why below the constraints from direct detection through dipoles will be competitive, despite the fact that they involve loop suppressed quantities.  

In Eqs.~\eqref{eq:m-23}-\eqref{eq:m-24}, $\alpha=1/137$, $Z$ is the atomic number, $F_{E}$ and $F_{M}$
are two nuclear form factors, and $G$ is a dimensionless quantity
depending on the nuclear spin $J$ and the nuclear magnetic dipole
$d_{N}$~\cite{Banks:2010eh}:
\begin{equation}
G=\frac{2(J+1)}{3J}\left(\frac{d_{N}A}{d_{n}}\right)^{2}\approx7256.78\ \text{(for Xe)}\thinspace,\label{eq:m-25}
\end{equation}
where $d_{n}=e/(2m_{p})$ is the nuclear magneton, and $A$ is the
mass number. For the nuclear form factors, we adopt the following
approximate expressions~\cite{Banks:2010eh}:
\begin{align}
F_{E} & =3\left[\frac{\sin(qr)-qr\cos(qr)}{(qr)^{3}}\right]e^{-q^{2}s^{2}}\thinspace,\label{eq:m-26}\\
F_{M} & =\frac{\sin(q\tilde{r})}{q\tilde{r}}\Theta(q\tilde{r}<2.55)+0.21\Theta(2.55<q\tilde{r}<4.5)\thinspace,\label{eq:m-27}
\end{align}
where $r=1.12A^{1/3}$ fm, $\tilde{r}=A^{1/3}$ fm, $s=1$ fm, and
$q=\sqrt{2E_{r}m_{N}}$. The $\Theta$ function takes either the value $1$
or the value $0$, depending on whether the condition it involves is satisfied. For
Xe targets, $q\tilde{r}<4.5$ corresponds to $E_{r}<117$ keV, which
in practice is always satisfied. 

In the magnetic dipole cross section \eqref{eq:m-23}
 we have included both SI ($\propto F_{E}^{2}$) and SD ($\propto F_{M}^{2}$)
parts because they can be equally important. For example, when $m_{\chi}=m_{N}$
and $v=1.2v_{\min}$, the ratio of the two parts at $E_{r}=30$ keV
is about $1.6$. For the electric dipole cross section \eqref{eq:m-24}
we have neglected a possible SD contribution because it is
highly suppressed. The fundamental reason for this is that electric
charges of nucleons can be added coherently, unlike magnetic moments of nucleons.   As a consequence, the
electric dipole cross section is generally much larger than the magnetic
one when $d_{M}\simeq d_{E}$.

\begin{figure}
\centering

\includegraphics[width=0.6\textwidth]{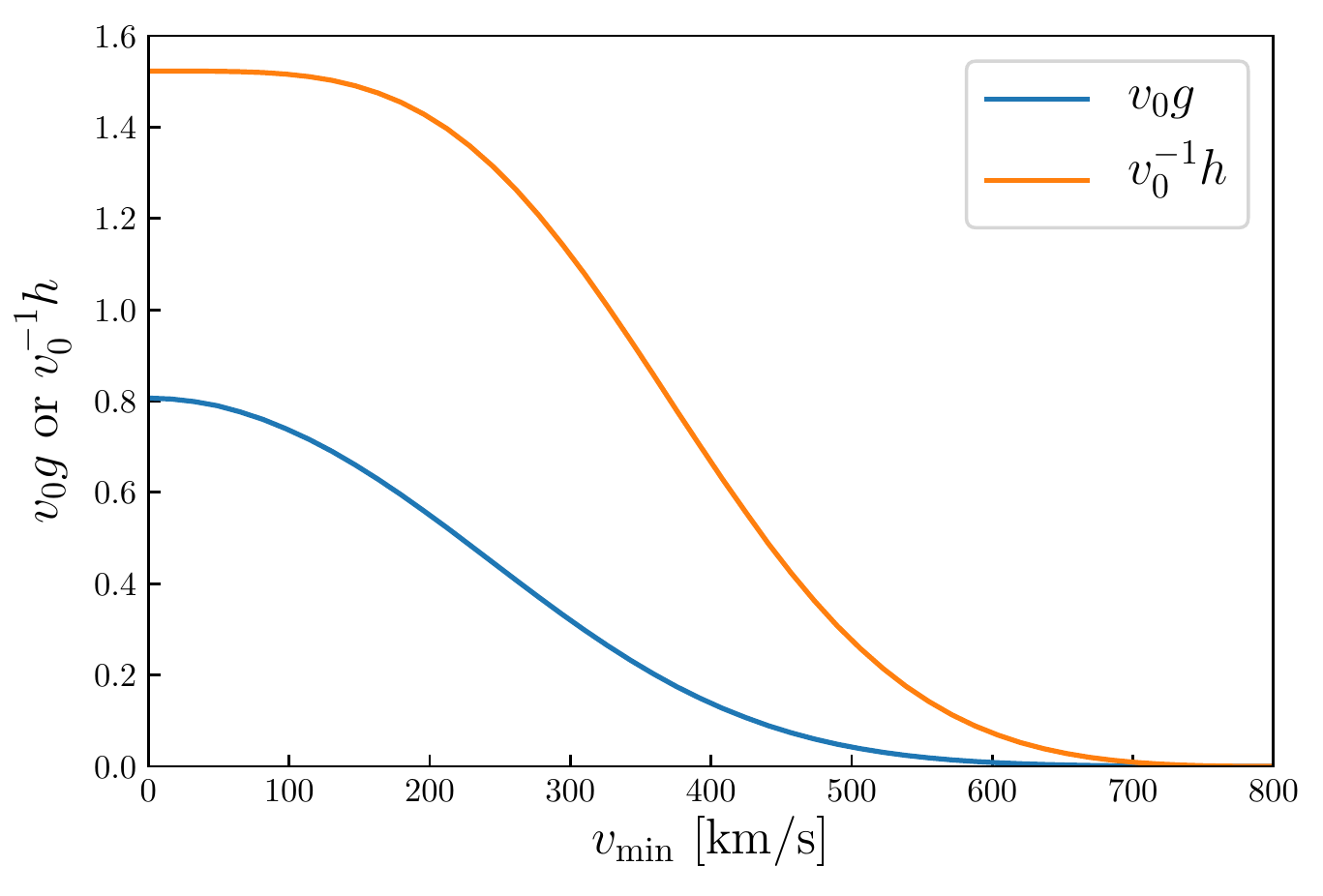}

\caption{\label{fig:gh}Results of $g(v_{\min})$ and $h(v_{\min})$ in Eqs.~\eqref{eq:m-30}
and \eqref{eq:m-31}. }
\end{figure}

For both Eqs.~\eqref{eq:m-23} and \eqref{eq:m-24}, the velocity
dependence can be written as follows:
\begin{equation}
\frac{d\sigma}{dE_{r}}=\frac{1}{v^{2}}\left(\frac{d\sigma_{g}}{dE_{r}}+v^{2}\frac{d\sigma_{h}}{dE_{r}}\right)\thinspace,\label{eq:m-28}
\end{equation}
where $d\sigma_{g}/dE_{r}$ and $d\sigma_{h}/dE_{r}$ are velocity
independent. Substituting Eq.~\eqref{eq:m-28} into Eq.~\eqref{eq:m-20},
we obtain
\begin{equation}
\frac{dR}{dE_{r}}=N_{T}n_{\chi}\epsilon(E_{r})\left[\frac{d\sigma_{g}}{dE_{r}}g(v_{\min})+\frac{d\sigma_{h}}{dE_{r}}h(v_{\min})\right]\thinspace,\label{eq:m-29}
\end{equation}
where 
\begin{align}
g(v_{\min}) & \equiv\int v^{-1}f_{\oplus}(\mathbf{v})\Theta(v-v_{\min})d^{3}\mathbf{v}\thinspace,\label{eq:m-30}\\
h(v_{\min}) & \equiv\int vf_{\oplus}(\mathbf{v})\Theta(v-v_{\min})d^{3}\mathbf{v}\thinspace,\label{eq:m-31}
\end{align}
can be computed independently of the cross section and of the kinematics of
DM-nucleus scattering. When numerically evaluating the integrals,
we take Eq.~\eqref{eq:m-21} with $\tilde{v}=(v^{2}+v_{\oplus}^{2}-2vv_{\oplus}\cos\theta)^{1/2}$
where $\theta$ is the angle between $\mathbf{v}_{\oplus}$ and $\mathbf{v}$,
and integrate $\theta$ from $0$ to $\pi$, $v$ from $0$ to $v_{0}+v_{{\rm esc}}$.
The results are presented in Fig.~\ref{fig:gh}.

\begin{figure}
    \centering
    
    \includegraphics[width=0.6\textwidth]{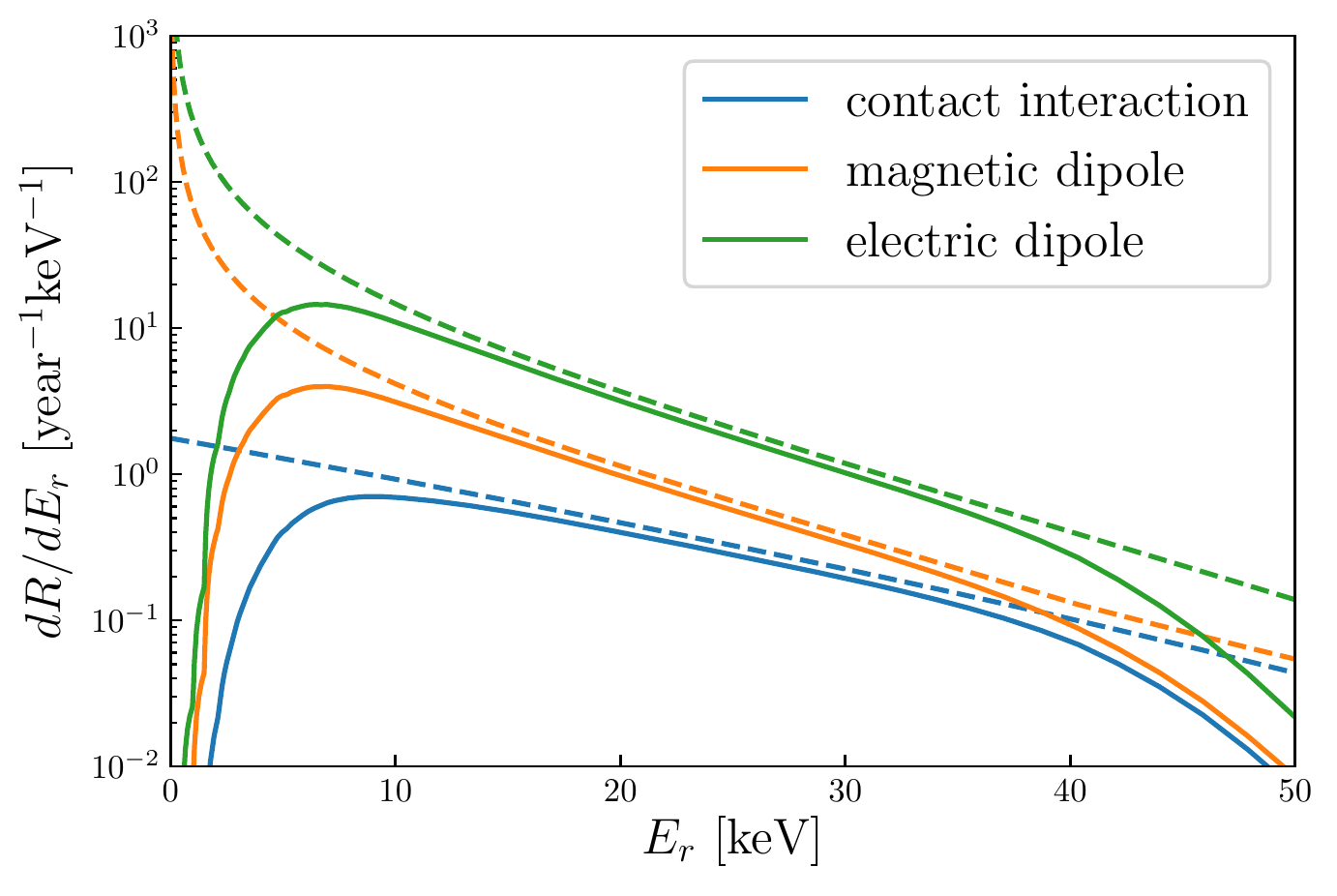}
    
    \caption{\label{fig:rate} Event rates for DM-nucleus scattering via magnetic
    and electric dipoles compared with the standard (contact-interaction)
    case. Shown examples take $d_{M}=5\times10^{-20}$ ecm, $d_{E}=1\times10^{-22}$
    ecm, $\sigma_{n}=10^{-46}\ \text{cm}^{2}$, and $m_{\chi}=100$ GeV.
    Dashed curves assume ideal detection efficiency 
    and solid curves take the XENON-1T detection efficiency~\cite{Aprile:2018dbl}
    into account. }
\end{figure}

Using Eq.~\eqref{eq:m-29}, we plot in Fig.~\ref{fig:rate} the differential
event rates (dashed curves) for $d_{M}=5\times10^{-20}$ ecm, $d_{E}=1\times10^{-22}$
ecm, assuming $m_{\chi}=100$ GeV and a $10^{3}$ kg liquid
Xe target. For comparison, we also present a curve for the following SI contact-interacting
cross section:
\begin{equation}
\frac{d\sigma_{{\rm contact}}}{dE_{r}}=\sigma_{n}\frac{m_{N}A^{2}F_{E}^{2}}{2\mu_{\chi n}^{2}v^{2}}\thinspace,\label{eq:m-32}
\end{equation}
where $\mu_{\chi n}=m_\chi m_n/(m_\chi+m_n)$
is the DM-nucleon reduced mass.
For the DM-nucleon cross section $\sigma_{n}$ we have taken the typical value that can be probed by direct detection experiment today, $\sigma_{n}=10^{-46}$~cm$^2$.
For the solid
curves in Fig.~\ref{fig:rate}, we have included the detection efficiency
of XENON-1T, which is taken from Fig.~1 in Ref.~\cite{Aprile:2018dbl}. 
The range of relevant recoil energy in direct detection experiments is relatively narrow, as below $\sim 5$~keV and above $\sim 50$~keV the efficiency is suppressed (compare dashed and solid lines in Fig.~\ref{fig:rate}). As can be seen in this figure too, within this range, the dipole lines display, as expected, an extra $1/E_r$ dependence with respect to the contact interaction case.

\begin{figure}
    \centering
    
    \includegraphics[width=0.49\textwidth]{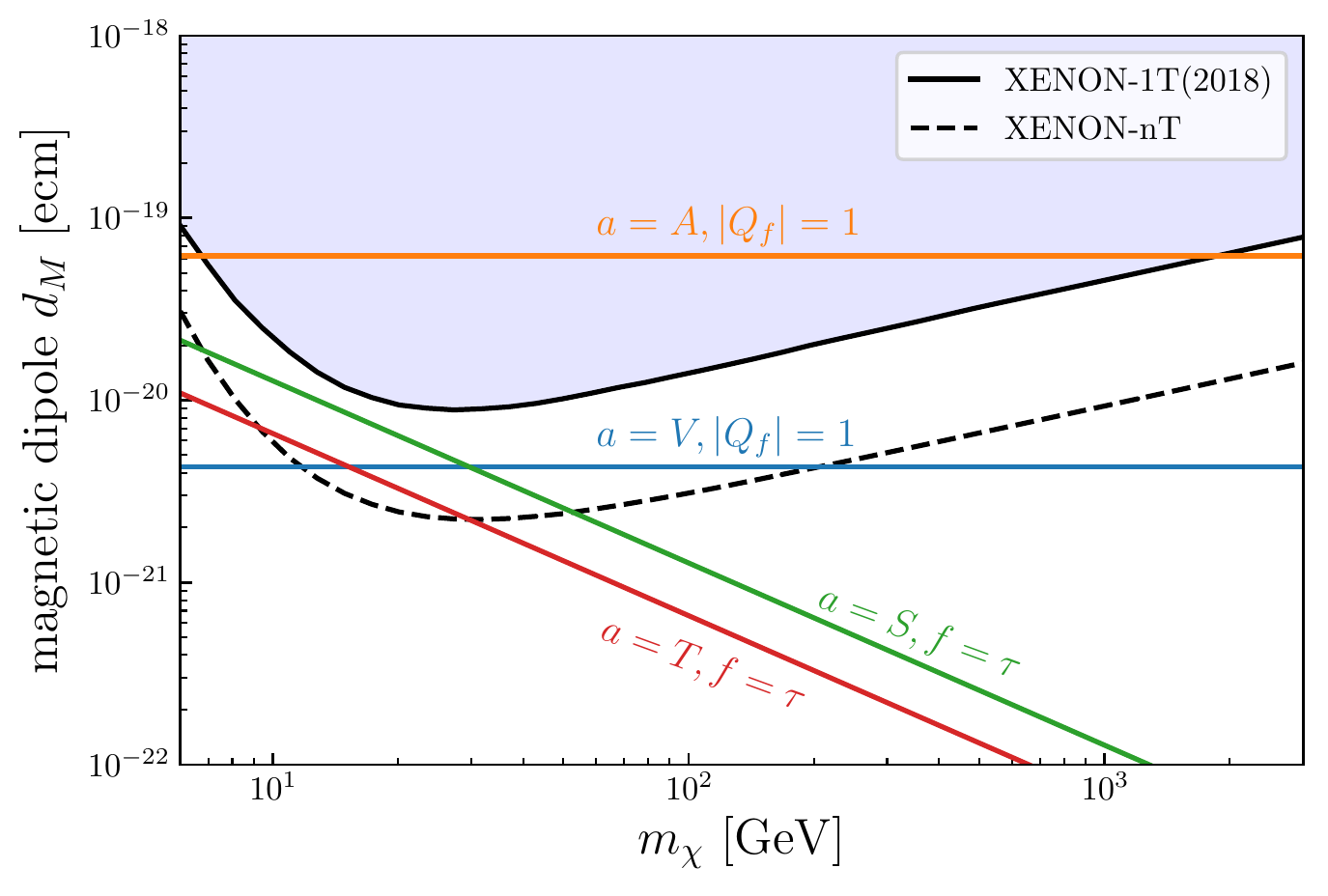}\includegraphics[width=0.49\textwidth]{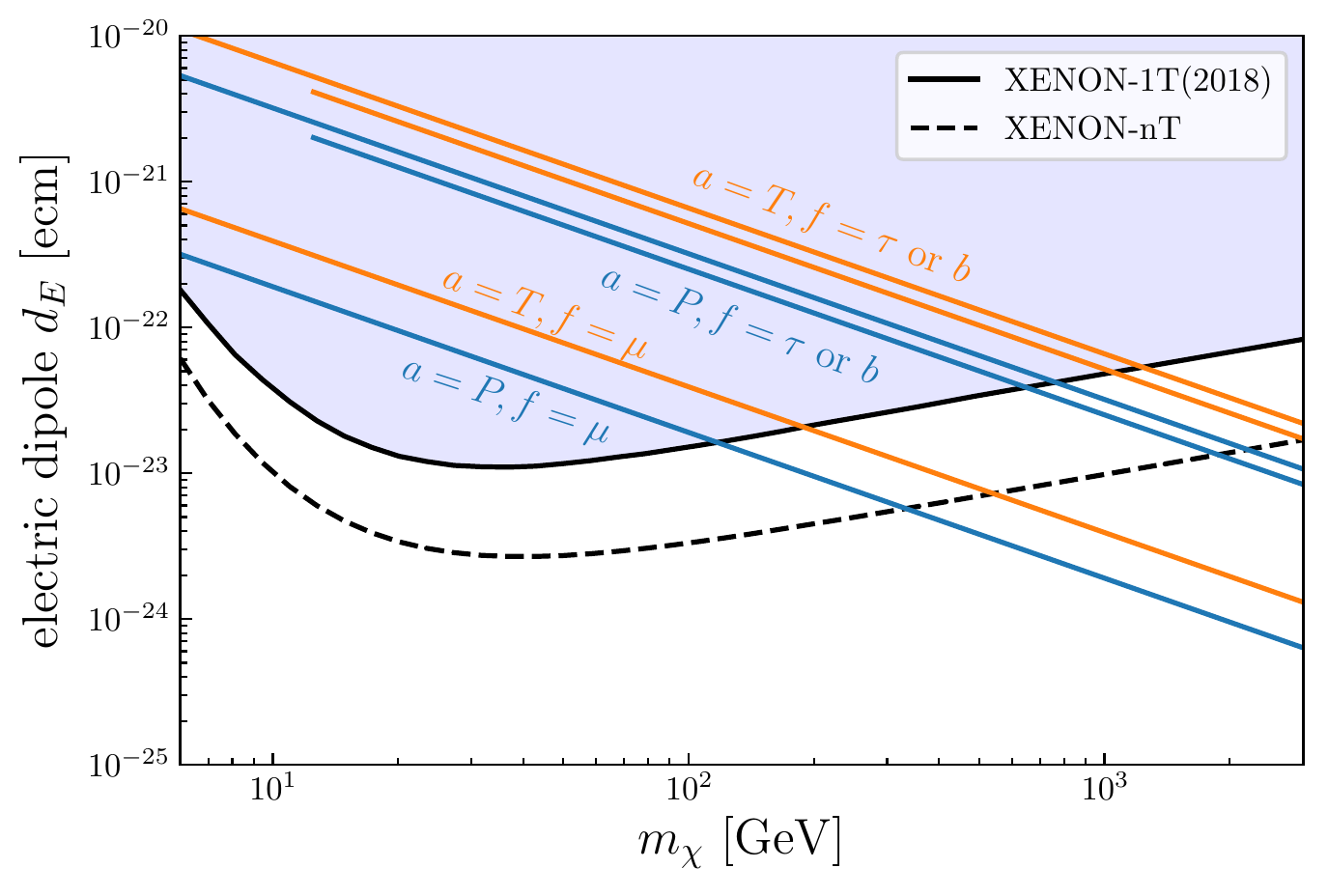}
    \caption{The loop-induced magnetic dipole $d_{M}$ (left panel) and electric
    dipole $d_{E}$ compared with the XENON-1T~\cite{Aprile:2018dbl}
    limit and the future XENON-nT~\cite{Aprile:2020vtw} sensitivity. The
    theoretical expectations of $d_{M}$ and $d_{E}$ are computed according
    to Eqs.~\eqref{eq:m-37}-\eqref{eq:m-40}. Dipoles for other fermions than the ones considered in these panels can be obtained for example from the $\tau$ case
    from a simple multiplication by $Q_f/Q_\tau$ for $V$ and $A$ and by $Q_f m_f/(Q_\tau m_\tau)$ for $S$, $P$ and $T$. \label{fig:final}}
    \end{figure} 
    
   \begin{figure}
     \centering
    \includegraphics[width=0.79\textwidth]{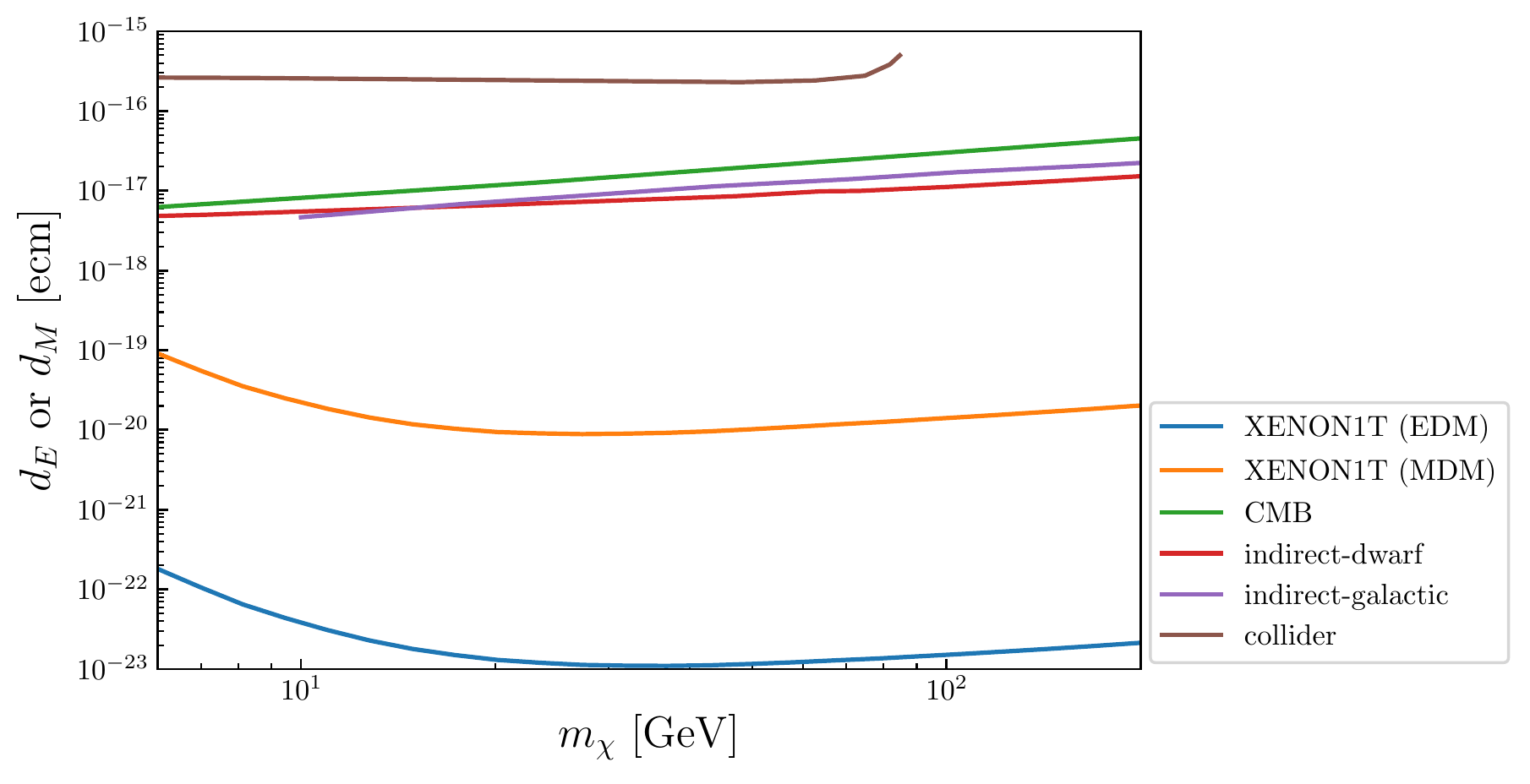}
    \caption{Comparison of direct detection bounds with other known bounds on DM electric/magnetic dipoles. The direct detection bounds (XENON-1T) are the same as those in Fig.~\ref{fig:final}.
    The CMB bound is taken from Ref.~\cite{DelNobile:2012tx}.
    The indirect detection bounds, also taken from Ref.~\cite{DelNobile:2012tx}, are derived from FERMI-LAT constraints on $\gamma$-rays from the Galaxy (labelled as indirect-galactic) and its dwarf satellite galaxies (indirect-dwarf). The collider bound is taken from Ref.~\cite{Fortin:2011hv}.
    \label{fig:all_bounds}}
    \end{figure}

As a result of this relatively narrow range of recoil energy, it is possible to recast the XENON-1T bounds obtained for a contact interaction into bounds holding for the massless mediator ($\propto 1/E_r$) case of interest here.
To this end, we apply a  spectrum-fitting
technique previously adopted in Ref.~\cite{Hambye:2018dpi}, namely
using Eq.~\eqref{eq:m-32} to fit the dipole-interacting recoil spectra.
Specifically, for a given set of $d_{M}$ ($d_{E}$) and $m_{\chi}$,
one can correspondingly find values of $\sigma_{n}$ and $m'_{\chi}$
(usually different from $m_{\chi}$) that minimizes the integral $\int\left(dR_{{\rm dipole}}/dE_{r}-dR_{\text{contact}}/dE_{r}\right)^{2}dE_{r}$
 where $dR_{{\rm dipole}}/dE_{r}$ and $dR_{\text{contact}}/dE_{r}$
are the dipole- and contact-interacting spectra (including the detection
efficiency). The minimization is performed under an additional constraint
that their total rates are equal. We find that after the minimization,
the two spectra are usually very close, with relative differences
typically below 20\%, which is consistent with the conclusion in Ref.~\cite{Hambye:2018dpi}.
By mapping $d_{M}$ ($d_{E}$)-$m_{\chi}$ to $\sigma_{n}$-$m'_{\chi}$
and taking the XENON-1T limit from Ref.~\cite{Aprile:2018dbl}, we
obtain the bounds on $d_{M}$ and $d_{E}$, presented in Fig.~\ref{fig:final}. 
For comparison, we also show  in Fig.~\ref{fig:all_bounds} other known bounds on DM electromagnetic dipoles from indirect detection, CMB observations, and collider searches. These bounds in the WIMP regime are known to be much weaker than that from direct detection.

Fig.~\ref{fig:final} shows that the possibility that the axial operators would be responsible for the observed DM relic density is already excluded by direct detection experiments within the $6.8~{\rm GeV}< m_{\chi}<1.9~{\rm TeV}$ range for charged leptons. 
Future experiments such as XENON-nT will enlarge this range significantly.
For the vector case, although it is beyond the current best limit from XENON-1T, future XENON-nT will be able to probe the range $11.8~{\rm GeV}< m_{\chi}<205~{\rm GeV}$.
The axial case is more constrained than the vector one because it requires a larger coefficient to account for the relic density constraint due to $p$-wave annihilation, see Eq.~\eqref{eq:m-7} and Tab.~\ref{tab:t}.
For the $S$, $P$ and $T$ cases the additional $m_f/m_\chi$ dependence of the dipoles decreases the sensitivity for high values of $m_\chi$ but boosts it for low values. The sensitivity also splits among generations of fermions. Taking the tensor case in the right panel of Fig.~\ref{fig:final} as an example, XENON-1T has excluded $m_{\chi}\lesssim 189$ GeV for $f=\mu$  while for $f=\tau$ this bound increases to $m_{\chi}\lesssim 1.2$ TeV.
The future experiment XENON-nT will be able to improve the mass bound by roughly a factor of three.

\begin{figure}
    \includegraphics[width=0.49\textwidth]{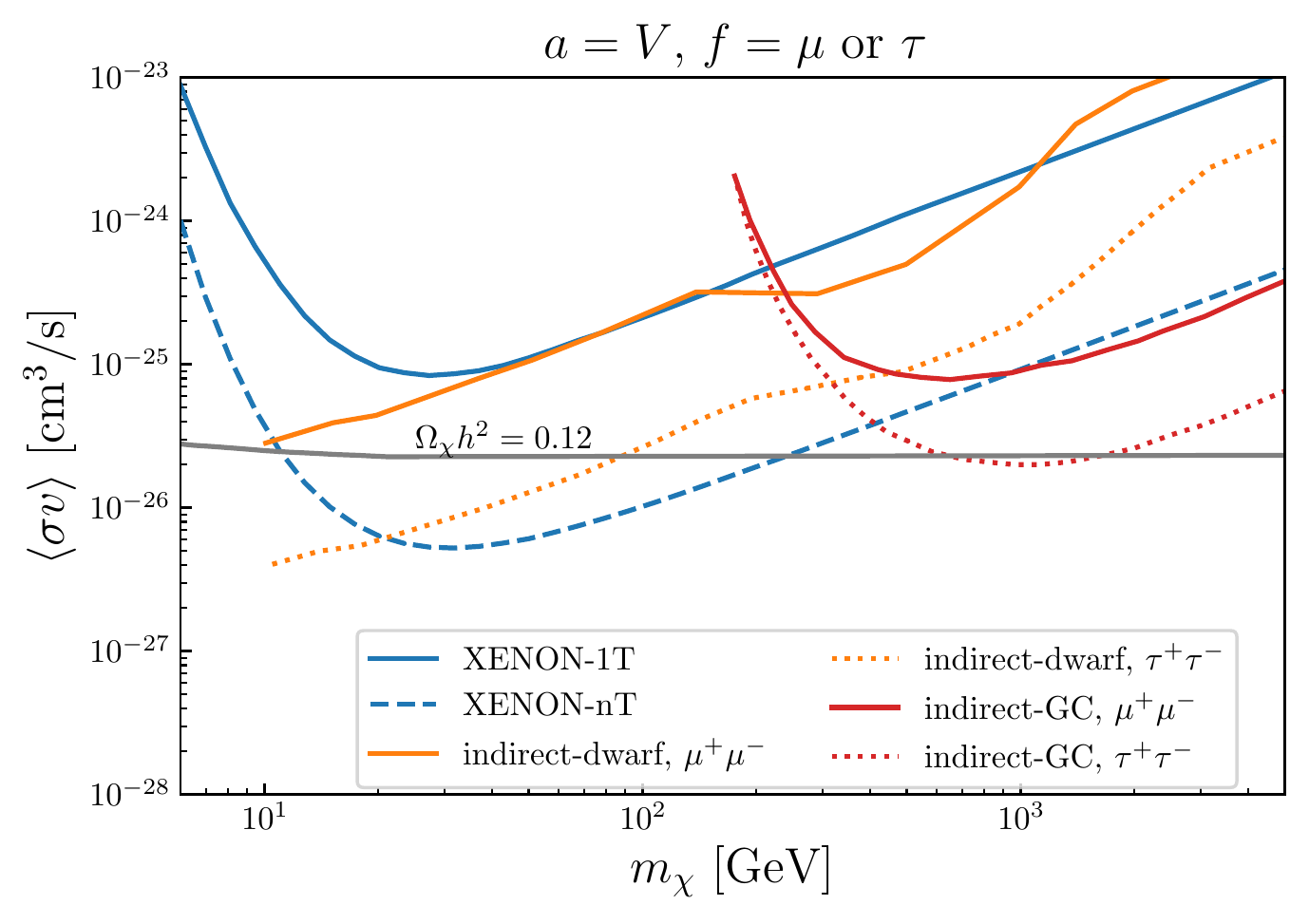}
    \includegraphics[width=0.49\textwidth]{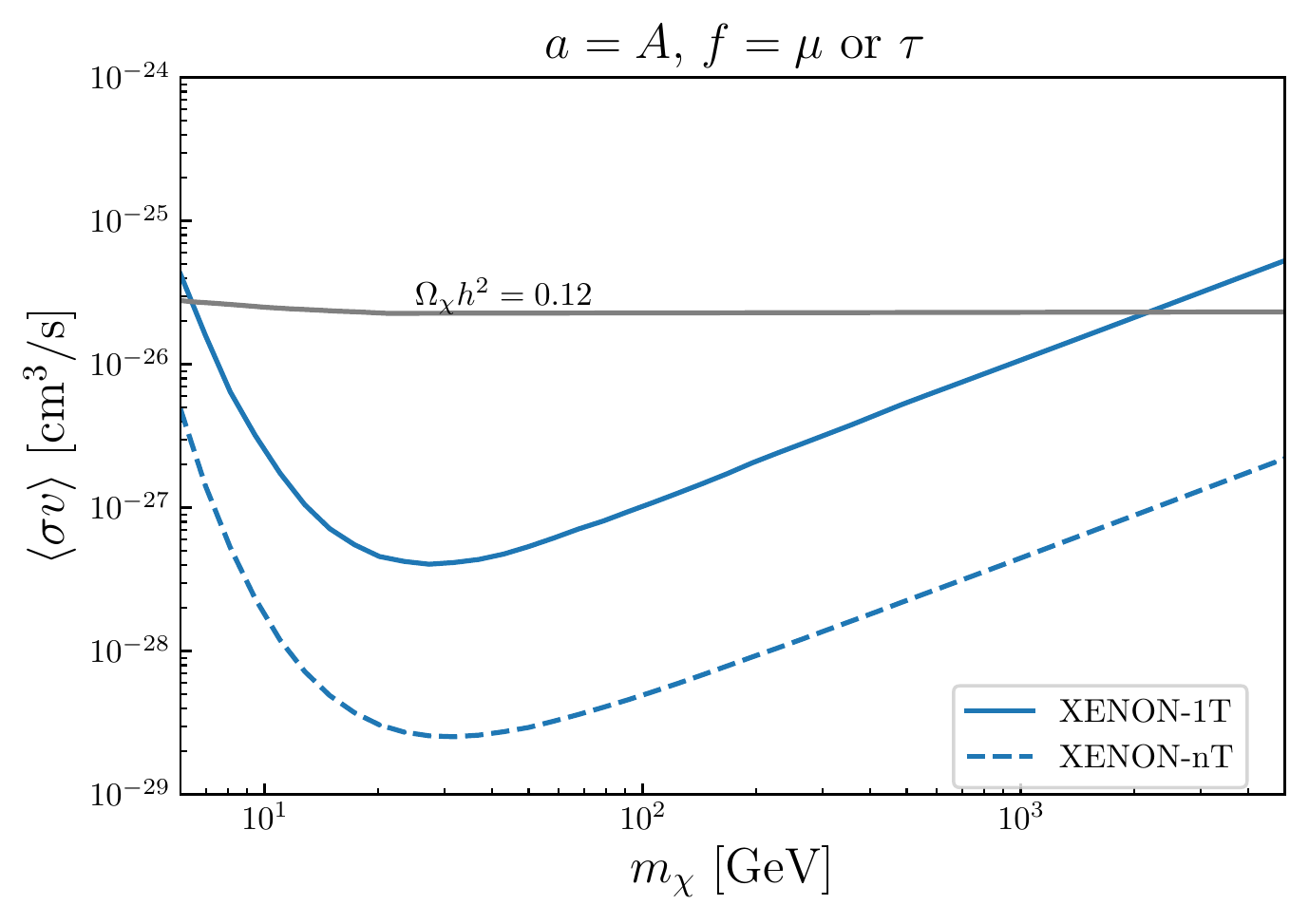}\\
    \includegraphics[width=0.49\textwidth]{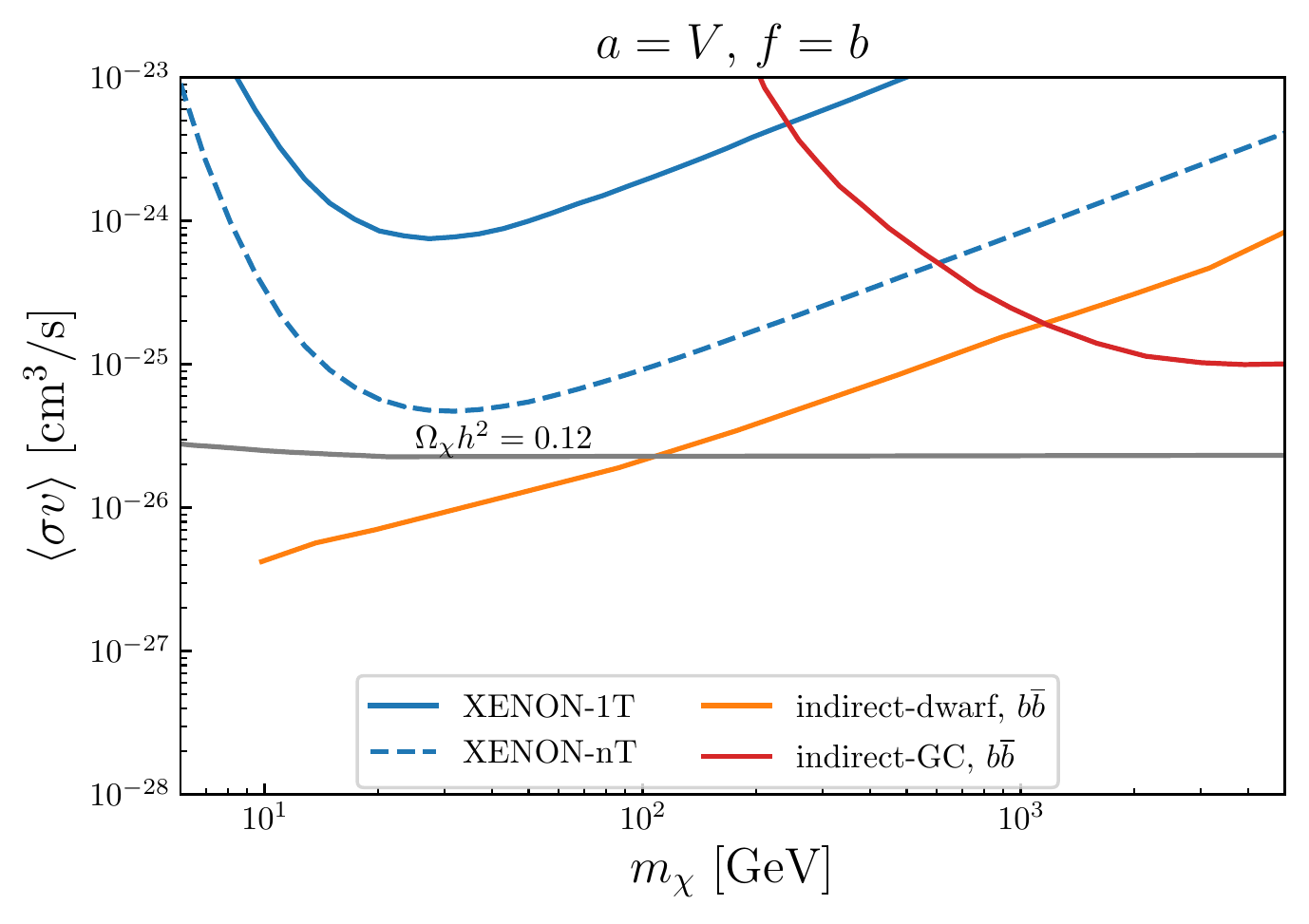}
    \includegraphics[width=0.49\textwidth]{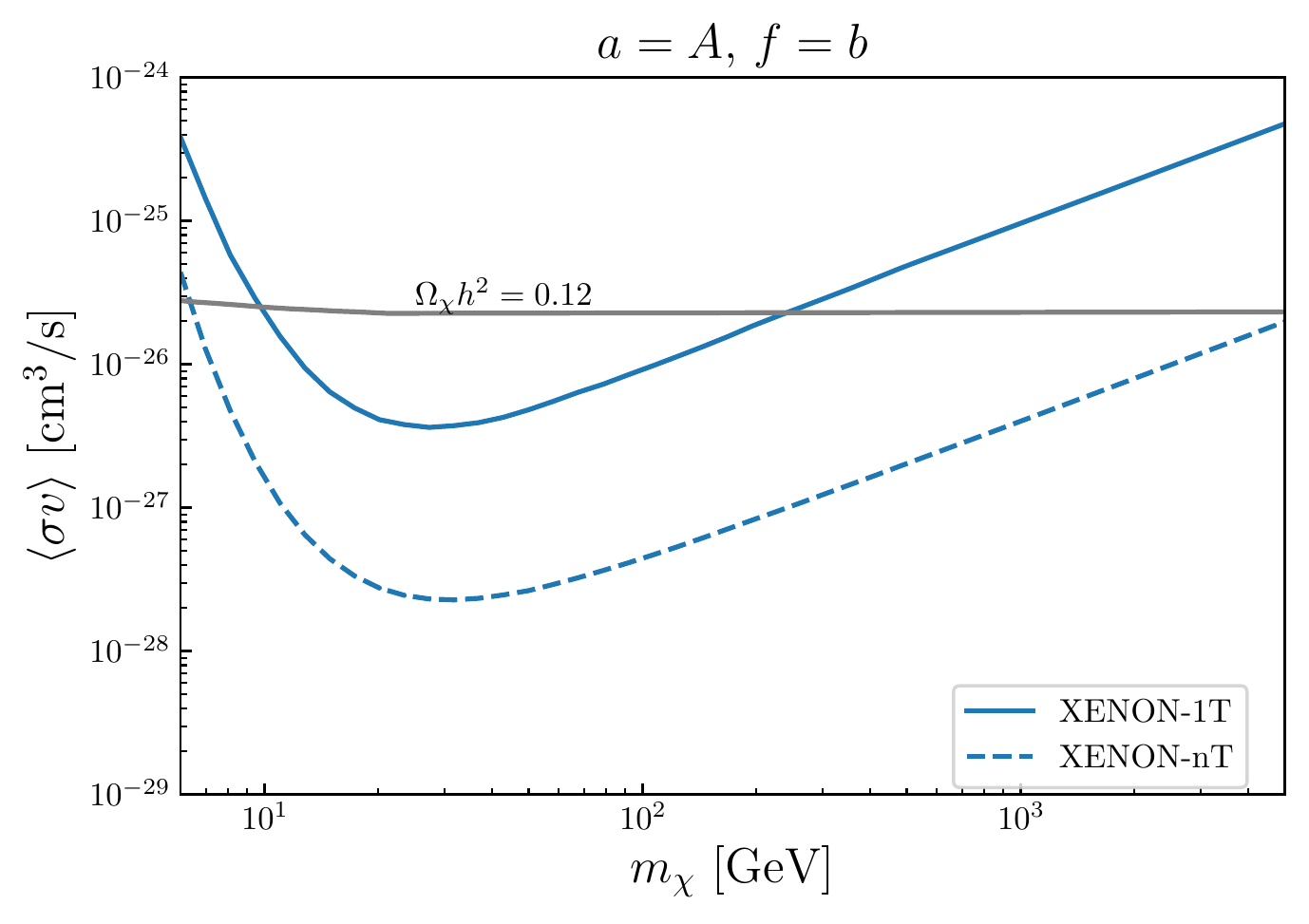}
    \caption{
    Direct detection bounds on $\langle\sigma v\rangle$ in comparison with indirect detection bounds, both obtained assuming $\Omega_{\chi} h^2=0.12$.
    For direct detection, we take the XENON-1T/nT bounds on $d_M$ obtained in Fig.~\ref{fig:final} and recast them to bounds on $\epsilon$'s according to Tab.~\ref{tab:dipole}, and further to  bounds on $\langle\sigma v \rangle$ according to Tab.~\ref{tab:t}, assuming $a=V$/$A$. For indirect detections, the ``indirect-dwarf'' bounds are taken from Ref.~\cite{Ahnen:2016qkx}, and   ``indirect-GC'' from \cite{Abdallah:2016ygi}, assuming that DM annihilates to $\tau^+\tau^-$, $\mu^+\mu^-$ or $b\overline{b}$.  The horizontal line shows the thermal cross section value.
    \label{fig:dir_indir}}
\end{figure}

\begin{figure}
    \centering
    \includegraphics[width=0.69\textwidth]{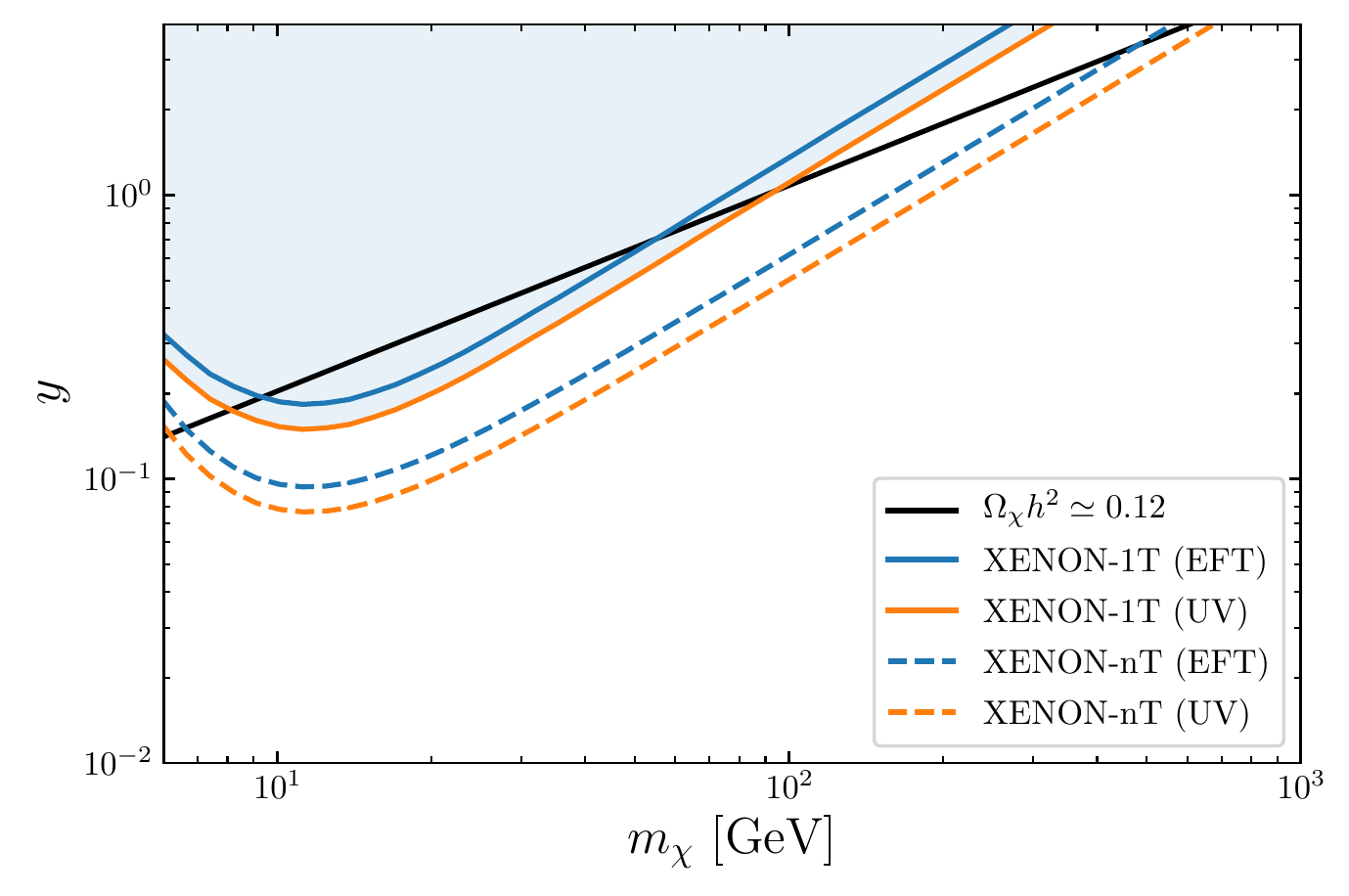} 
    \caption{Constraints on the UV complete model defined in Eq.~\eqref{eq:x-8} for any charged lepton $f$ and $m_{\phi}=2m_{\chi}$. EFT/UV indicates that the curve takes the loop-induced dipole in Eq.~\eqref{eq:x-14}/\eqref{eq:dMUV}.
    \label{fig:model}}
    \end{figure}

The four fermion interactions also induce fluxes of cosmic rays from the annihilation into charged fermion they induce today at tree level in the galactic center and dwarf galaxies. Indirect detection experiments give upper bounds on these fluxes which are generally translated into upper bounds on the annihilation cross section assuming $\Omega_{\chi} h^2=0.12$ (i.e.~not looking at the implications that this annihilation cross section could have on the relic density). In Fig.~\ref{fig:dir_indir} we show for the $V$ case how these bounds compare with the bounds that can be obtained on the same cross sections from the bound that direct detection set on the dipoles and thus on the coefficient of the four-fermion operators (assuming $\Omega_{\chi} h^2=0.12$ anyway there too).
As Fig.~\ref{fig:dir_indir} shows, despite that at tree level direct detection experiments are not much sensitive to the four-fermion operators for charged leptons or heavy quarks, when the loop-induced dipoles are taken into account,  direct detection offers competitive constraints on such operators in comparison with indirect detection. For the $A$ case (and similarly for the $S$ case)
indirect detection constraints are known to be weak as in this case the annihilation is of the $p$-wave type. But the bounds from dipole induced direct detection remains fully relevant, as given in Fig.~\ref{fig:dir_indir} too.

Explicit UV complete models can be constrained according to the combination of effective operators they lead to. 
For the model considered in Sec.~\ref{sub:example}, which in a characteristic way gives both $V$ and $A$ interactions with similar weights, we show in Fig.~\ref{fig:model} upper bounds on the Yukawa coupling $y$ of Eq.~(\ref{eq:x-8}), using the dipoles obtained in Eq.~\eqref{eq:x-14} (EFT) and Eq.~\eqref{eq:dMUV} (UV).  
As previously discussed in Sec.~\ref{sub:example}, the difference between the two approaches shown in Fig.~\ref{fig:model}  is small.
The current XENON-1T limit excludes the mass range $8.1~{\rm GeV} \lesssim m_{\chi}\lesssim 94~{\rm GeV}$ 
while future XENON-nT will be able to probe the $6.2~{\rm GeV} \lesssim m_{\chi}\lesssim 843~{\rm GeV}$ range.

As mentioned in the introduction, Ref.~\cite{Kopp:2009et} also studied the direct detection signals that could be induced by four-fermion effective operators (involving two charged leptons) via DM-photon interactions induced at the one-loop level. This calculation was done by calculating the loop directly from the EFT and applying an $\overline{\hbox{MS}}$ prescription. Before concluding let us add a few comments on the several improvements we made here. Beside the fact that, strictly speaking, the divergences obtained at the pure EFT level cannot be removed applying this prescription (since there is no counterterm that could 
cancel
the divergences, i.e.~the result must be finite), opening the four-fermion blob as we did above shows that the results can largely depend on the way these effective interactions are generated (as shown by the $s$-channel example which gives vanishing results). The two-step calculation we did, calculating first the various moments that are induced and subsequently computing what it gives for direct detection (instead of calculating directly the direct detection cross section) is useful 
as it identifies for each case 
what kind of electromagnetic interactions is induced 
(with what it implies for each of these interactions). 
Phenomenologically Fig.~\ref{fig:final} also shows that 
the dramatic improvements of direct detection experiments in the last decade imply 
that the cases with chirality-flip suppressions (i.e.~the $S$, $P$ and $T$ cases)
are also testable (see e.g.~the scalar case for $f=\tau$ in Fig.~\ref{fig:final}).
As also shown above, electromagnetic interactions induced at the one-loop level are relevant not only for charged leptons but also for the heavy quark case.\footnote{For this case, it would be interesting to compute the loop-level direct detection signals that are induced via gluon exchange rather than photon exchange, which is beyond the scope of this work.}

\section{Conclusion\label{sec:Conclusion}}
\noindent
In the presence of four-fermion effective interactions
of dark matter (DM) with Standard Model (SM) fermions, electromagnetic
dipoles of DM can easily be generated, due to the loop process illustrated in Fig.~\ref{fig:close-f}. This is the case in particular if the operators are generated through the exchange of a $t$-channel mediator.
We study systematically for all possible effective interactions the loop-induced dipoles and find that, if they are not identically vanishing,
the electromagnetic dipoles in the WIMP paradigm are typically of the order of $10^{-21}$ ($10^{-20}$) ecm for vector (axial-vector) interactions, or of $10^{-20} m_f/m_{\chi}$ ecm for scalar, pseudo-scalar, and tensor interactions, see Eqs.~\eqref{eq:m-37}-\eqref{eq:m-40}. Calculations for a UV complete model give very similar results.

Via photon exchange such values imply observable nuclear recoil signals in direct detection experiments. 
This provides (or will provide) the most stringent constraints for various operators, in particular for axial or scalar operators, as well as for operators involving for instance muons.
So far XENON-1T has excluded the loop-induced electromagnetic dipoles for some types of effective interactions in certain mass ranges---see Fig.~\ref{fig:final}. Future multi-ton liquid xenon experiments with substantially improved sensitivity will be able to probe the dipoles for all types of effective interactions over much broader mass ranges.

\begin{acknowledgments}
\noindent
We thank Xiaoyong Chu and Laurent Vanderheyden for useful discussions.
This work is supported by the ``Probing dark matter with neutrinos'' ULB-ARC convention and by the F.R.S./FNRS under the Excellence of Science (EoS) project No.\ 30820817 - be.h ``The $H$ boson gateway to physics beyond the Standard Model''.
\end{acknowledgments}

\appendix

\section{Loop calculations\label{sec:loop-cal}}
\noindent
In this appendix, we present the details of our loop calculations.
Starting from Eq.~\eqref{eq:m-9}, we first compute the traces:
\begin{align}
{\rm tr}\left[\frac{1}{\slashed{k}_{2}-m_{f}}\gamma^{\mu}\frac{1}{\slashed{k}_{1}-m_{f}}P_{L}\right] & =2m_{f}(k_{1}^{\mu}+k_{2}^{\mu})D_{k}\thinspace,\label{eq:m-11}\\
{\rm tr}\left[\frac{1}{\slashed{k}_{2}-m_{f}}\gamma^{\mu}\frac{1}{\slashed{k}_{1}-m_{f}}\gamma^{\nu}P_{L}\right] & =2\left[k_{1}^{\mu}k_{2}^{\nu}+k_{2}^{\mu}k_{1}^{\nu}+(m_{f}^{2}-k_{2}\cdot k_{1})g^{\mu\nu}-i\epsilon^{\mu\nu\rho\lambda}k_{1\rho}k_{2\lambda}\right]D_{k}\thinspace,\label{eq:m-12}\\
{\rm tr}\left[\frac{1}{\slashed{k}_{2}-m_{f}}\gamma^{\mu}\frac{1}{\slashed{k}_{1}-m_{f}}\sigma^{\rho\lambda}P_{L}\right] & =2im_{f}\left[(k_{1}^{\rho}-k_{2}^{\rho})g^{\lambda\mu}-\lambda\leftrightarrow\rho\right]+2m_{f}\epsilon^{\lambda\mu\rho\nu}(k_{1}-k_{2})_{\nu}D_{k}\thinspace,\label{eq:m-13}
\end{align}
where 
\begin{equation}
D_{k}\equiv\frac{1}{k_{2}^{2}-m_{f}^{2}}\frac{1}{k_{1}^{2}-m_{f}^{2}}\thinspace.\label{eq:x-4}
\end{equation}
If $P_{L}$ in the above traces is replaced by $P_{R}$, the results
are similar except that $\epsilon^{\mu\nu\rho\lambda}$ and $\epsilon^{\lambda\mu\rho\nu}$
flip their signs. 

Next, we plug the traces into the loop integral and integrate out
$k$, assuming the mass hierarchy: 
\begin{equation}
m_{\text{med}}\gg m_{\chi}\gg m_{f}.\label{eq:x-7}
\end{equation}
Taking the case of Eq.~\eqref{eq:m-11} with $G_{X}^{(t)}$ in Eq.~\eqref{eq:m-14}
for example, we have:
\begin{equation}
\int\frac{d^{4}k}{(2\pi)^{4}}{\rm tr}\left[\frac{1}{\slashed{k}_{2}-m_{f}}\gamma^{\mu}\frac{1}{\slashed{k}_{1}-m_{f}}P_{L}\right]\frac{1}{k^{2}-m_{\text{med}}^{2}}\approx-\frac{iy^{2}}{16\pi^{2}}\left(p_{1}+p_{2}\right)^{\mu}\frac{m_{f}}{m_{\text{med}}^{2}}\thinspace.\label{eq:x-5}
\end{equation}
The loop integral is computed using {\tt Package-X}~\cite{Patel:2015tea}
and expanded in $q^{2}$, $m_{f}$, and $m_{\chi}$. Only the leading
order term is taken. Note that the integral is free from UV divergences
because for $k\rightarrow\infty$ the integral behaves like $\int k^{-5}d^{4}k$. 

With the result in Eq.~\eqref{eq:x-5} and $y^{2}/m_{\text{med}}^{2}=\epsilon_{a}^{L}G_{F}$,
we see that ${\cal F}^{\mu}$ introduced in Eq.~\eqref{eq:m-9} for
$a=S$ and $P$ should be
\begin{equation}
{\cal F}^{\mu}=\Gamma^{a}eQ_{f}\frac{\epsilon_{a}^{L}G_{F}m_{f}}{16\pi^{2}}\left(p_{1}+p_{2}\right)^{\mu}.\label{eq:x-6}
\end{equation}
Then for $\overline{u_{2}}{\cal F}^{\mu}u_{1}$, it can be decomposed
into the form factors in Eq.~\eqref{eq:m-10} using the following
 identities: 
\begin{align}
\overline{u_{2}}(p_{1}+p_{2})^{\mu}u_{1} & =\overline{u_{2}}\left[i\sigma^{\mu\nu}q_{\nu}\right]u_{1}+2m_{\chi}\overline{u_{2}}\gamma^{\mu}u_{1}\label{eq:x}\\
\overline{u_{2}}(p_{1}+p_{2})^{\mu}i\gamma^{5}u_{1} & =\overline{u_{2}}\left[-\sigma^{\mu\nu}\gamma^{5}q_{\nu}\right]u_{1},\label{eq:x-1}
\end{align}
where $q\equiv p_{1}-p_{2}$. Eq.~\eqref{eq:x} is the well-known
Gordon identity (due to the definition of $q$, our convention differs
from that in Ref.~\cite{Peskin} by a minus sign of $q$).  Eq.~\eqref{eq:x-1}
is similar but with additional $\gamma^{5}$. It can be derived as
follows: 
\begin{align}
\overline{u_{2}}\left[i\sigma^{\mu\nu}\gamma^{5}q_{\nu}\right]u_{1} & =-\frac{1}{2}\overline{u_{2}}[\gamma^{\mu},\gamma^{\nu}]\gamma^{5}q_{\nu}u_{1}\nonumber \\
 & =-\frac{1}{2}\overline{u_{2}}\left[\gamma^{\mu}\slashed{p}_{1}\gamma^{5}-\gamma^{\mu}\slashed{p}_{2}\gamma^{5}-\slashed{p}_{1}\gamma^{\mu}\gamma^{5}+\slashed{p}_{2}\gamma^{\mu}\gamma^{5}\right]u_{1}\nonumber \\
 & =\frac{1}{2}\overline{u_{2}}\left[(2p_{2}^{\mu}-\slashed{p}_{2}\gamma^{\mu})\gamma^{5}+(2p_{1}^{\mu}-\gamma^{\mu}\slashed{p}_{1})\gamma^{5}\right]u_{1}\nonumber \\
 & =(p_{1}+p_{2})^{\mu}\overline{u_{2}}\gamma^{5}u_{1},\label{eq:x-2}
\end{align}
where in the second row the first and last terms cancel out because
$\slashed{p}_{1}\gamma^{5}u_{1}=-m_{\chi}\gamma^{5}u_{1}$ and $\overline{u_{2}}\slashed{p}_{2}=\overline{u_{2}}m_{\chi}$,
and in the third row the $\slashed{p}_{2}\gamma^{\mu}$ and $\gamma^{\mu}\slashed{p}_{1}$
terms cancel out for the same reason. 

According to Eqs.~\eqref{eq:x} and \eqref{eq:x-1}, Eq.~\eqref{eq:x-6}
generates magnetic and electric dipoles for $a=S$ and $P$, respectively.
The values are already listed in Tab.~\ref{tab:dipole}.  For other
cases, the calculations are similar: We plug Eq.~\eqref{eq:m-12}
or \eqref{eq:m-13} into the loop integral to obtain ${\cal F}^{\mu}$
and use Eq.~\eqref{eq:x} or \eqref{eq:x-1} to extract the dipole
form factors. In general, when the resulting ${\cal F}^{\mu}$ contains
$\slashed{p}_{1}$ and $\slashed{p}_{2}$, after applying the Dirac
algebra and on-shell conditions ($\slashed{p}_{1}u_{1}=m_{\chi}u_{1}$
and $\overline{u_{2}}\slashed{p}_{2}=\overline{u_{2}}m_{\chi}$),
they can be converted to linear combinations of $(p_{1}+p_{2})^{\mu}$
and $(p_{1}-p_{2})^{\mu}$. Terms containing the latter cancel out
or can be neglected due to the Ward identity. In {\tt Package-X}~\cite{Patel:2015tea},
the dipole form factors can be extracted using dedicated projectors,
and we have verified that this approach leads to the same results. 



As for the UV complete example introduced in Sec.~\ref{sub:example}, there are two diagrams contributing to the magnetic dipole: one with the photon coupled to the charge fermion $f$ and the other with photon coupled to the charged scalar $\phi^{\pm}$. We refer to the former and the latter
as diagrams (i) and (ii), respectively. Their amplitudes read
\begin{align}
i{\cal M}_{{\rm (i)}} & =i\int\frac{d^{4}k}{(2\pi)^{4}}\overline{u_{2}}P_R\frac{1}{\slashed{k}_{2}-m_{f}}ieQ_{f}\gamma^{\mu}\frac{1}{\slashed{k}_{1}-m_{f}}P_L u_{1}\varepsilon_{\mu}\frac{yy^{*}}{k^{2}-m_{\phi}^{2}}\thinspace,\nonumber \\
i{\cal M}_{{\rm (ii)}} & =i\int\frac{d^{4}k}{(2\pi)^{4}}\overline{u_{2}}P_R \frac{1}{\slashed{k}-m_{f}}P_L u_{1}\varepsilon_{\mu}ieQ_{\phi}(k_{1}+k_{2})^{\mu}\frac{1}{k_{1}^{2}-m_{\phi}^{2}}\frac{1}{k_{2}^{2}-m_{\phi}^{2}}yy^{*}\thinspace,\label{eq:m-9-2}
\end{align}
where $Q_{\phi}$ is the electric charge of $\phi^{\pm}$. 
After integrating out the loop momentum, we find
\begin{align}
    {\cal M}_{{\rm (i)}} & =\frac{e|y|^{2}Q_{f}}{32\pi^{2}}\overline{u_{2}}\gamma^{\mu}u_{1}\varepsilon_{\mu}\left[\frac{1}{2\epsilon}+\frac{1}{4}+\log\left(\frac{\mu}{m_{\phi}}\right)\right]\nonumber \\
     & +\frac{e|y|^{2}Q_{f}}{32\pi^{2}m_{\phi}^{2}}\overline{u_{2}}\gamma^{\mu}u_{1}\varepsilon_{\mu}\left[\frac{1}{2}m_{\chi}^{2}-\frac{1}{2}m_{f}^{2}\right]\nonumber \\
     & +\frac{e|y|^{2}Q_{f}}{32\pi^{2}m_{\phi}^{2}}\overline{u_{2}}i\sigma^{\mu\nu}q_{\nu}u_{1}\varepsilon_{\mu}\left[-\frac{1}{3}m_{\chi}\right]\nonumber \\
     & +\left[\overline{u_{2}}\gamma^{\mu}\gamma^{5}u_{1}\ {\rm terms}\right],\label{eq:x-11}\\
    {\cal M}_{{\rm (ii)}} & =\frac{e|y|^{2}Q_{\phi}}{32\pi^{2}}\overline{u_{2}}\gamma^{\mu}u_{1}\varepsilon_{\mu}\left[\frac{1}{2\epsilon}+\frac{1}{4}+\log\left(\frac{\mu}{m_{\phi}}\right)\right]\nonumber \\
     & +\frac{e|y|^{2}Q_{\phi}}{32\pi^{2}m_{\phi}^{2}}\overline{u_{2}}\gamma^{\mu}u_{1}\varepsilon_{\mu}\left[\frac{1}{2}m_{\chi}^{2}-\frac{1}{2}m_{f}^{2}\right]\nonumber \\
     & +\frac{e|y|^{2}Q_{\phi}}{32\pi^{2}m_{\phi}^{2}}\overline{u_{2}}i\sigma^{\mu\nu}q_{\nu}u_{1}\varepsilon_{\mu}\left[+\frac{1}{6}m_{\chi}\right]\nonumber \\
     & +\left[\overline{u_{2}}\gamma^{\mu}\gamma^{5}u_{1}\ {\rm terms}\right].\label{eq:x-12}
    \end{align}
    Here $\overline{u_{2}}\gamma^{\mu}\gamma^{5}u_{1}$ terms are not
    important because they cancel out in the final result, as we have
    verified explicitly in the calculation.

As is manifest, the UV divergences in the first rows of Eqs.~\eqref{eq:x-11}
and \eqref{eq:x-12} cancel out when $Q_{\phi}+Q_{f}=0$, which is
required by the charge conservation of Eq.~\eqref{eq:x-8}. What also
cancel out in the remaining terms are those proportional to $\overline{u_{2}}\gamma^{\mu}u_{1}$,
which is expected because $\chi$ is neutral. 

After all the cancellations, only the magnetic dipole terms exist.
Comparing the last row of Eq.~\eqref{eq:x-11} to Eq.~\eqref{eq:x-14},
we see that ${\cal M}_{{\rm (i)}}$ reproduces the dipole obtained in the EFT approach.
When the full theory is taken into account,
the additional contribution due to ${\cal M}_{{\rm (ii)}}$  is roughly
half the size of the previous one, assuming $m_{f}\ll m_{\chi}$. 
Taking $Q_{\phi}=-Q_f$ and summing the two diagrams together, we obtain the result in Eq.~\eqref{eq:dMUV}.

\bibliographystyle{JHEP}
\bibliography{ref}

\end{document}